\documentclass[12pt]{revtex4-1}
\usepackage{graphicx}
\usepackage{amsfonts}
\usepackage{amsmath}
\usepackage{amssymb}
\renewcommand\O{\mathcal{O}}
\newcommand\sep{\vspace{.5cm}\noindent}
\newcommand\lift{\vspace{-.2cm}}
\newcommand\D[2]{\left.\frac{\partial}{\partial{#1}^{#2}}\right|_p}
\newcommand\DD[2]{\frac{\partial}{\partial{#1}^{#2}}}
\newcommand\new[1]{#1}
\newcommand\newpageja{}
\renewcommand\min{-}

\begin{document}
\enlargethispage{2cm}
\title{Obergurgl Lectures on\\[6pt] ALL SPACETIMES BEYOND EINSTEIN} 
\author{\vspace{.6cm}Frederic P. Schuller}
\address{\vspace{.2cm}Albert-Einstein-Institut\\ Max-Planck-Institut f\"ur Gravitationsphysik\\ Am M\"uhlenberg 1, 14476 Potsdam, Germany\\ fps@aei.mpg.de}
\begin{abstract}
Which geometries on a smooth manifold (apart from Lorentzian metrics) can serve as a spacetime structure?
This question is comprehensively addressed from first principles in eight lectures,  exploring the kinematics and gravitational dynamics of all tensorial geometries on a smooth manifold that can carry predictive matter equations, are time-orientable, and allow to distinguish positive from negative particle energies.
\end{abstract}
\maketitle

\section*{INTRODUCTION}
The recent announcement of superluminal neutrino propagation by the OPERA collaboration is a reminder that the spacetime geometry might well not be given by a Lorentzian manifold. But by what else?
Fortunately, this question has a rather comprehensive answer. For we will show in these lectures that the spectrum of tensor fields that can serve as a spacetime geometry---in the sense that matter field dynamics is predictive and observers agree on the sign of particle energies---is severely restricted. Only geometries on which the dispersion relation of matter fields is encoded in a totally symmetric contravariant even-rank tensor field $P$ satisfying three simple algebraic conditions -- it must be hyperbolic, time-orientable and energy-distinguishing -- can present candidates for a spacetime geometry. These physically inevitable properties single-handedly ensure that the entire kinematical apparatus familiar from physics on a Lorentzian manifold is defined precisely the same way for any spacetime; causality, in particular, is perfectly compatible with superluminal propagation in spacetimes, but one only learns this from a subtle interplay of convex analysis, real algebraic geometry and the modern theory of partial differential equations. The relevance of this class of geometries roots in that fact that they already exhaust the possible spacetime structures on a smooth manifold. Thus this class, and only this class of geometries, merits our attention.

Lorentzian manifolds simply present the simplest example of a tensorial spacetime geometry. But considering any one of the more refined geometries of course eradicates the very foundation of Einstein's field equations as we know them. It does not eradicate, though, the deeper principle behind them that was revealed by the Wheeler school a long time ago: in their geometrodynamical view, gravitational dynamics is all about evolving the spatial geometry from one  suitable initial data surface to an infinitesimally neighbouring one, such that ultimately all spatial geometries recombine to an admissible spacetime geometry. The geometrodynamic principle stands independent of the particular spacetime geometry it is applied to. Thus the second key insight arrived at in these lectures is that finding the gravitational dynamics of refined spacetime geometries does not require inspired physical guesswork, but is reduced to solving a clear-cut mathematical problem. It remains of course for experiment to decide which member of the only countable class of classical spacetime theories is realized in Nature.
 
\newpage
\section*{Lecture I: \, MANIFOLDS}
Every study needs to start from foundations that are not further questioned. For the present lectures, this is the assumption that spacetime is certainly a smooth $d$-dimensional manifold $M$. For the benefit of the non-specialist, this first lecture recalls the relevant definitions from topology and differentiable manifold theory.

\sep{\bf Definition.} A set $M$ is made into a \underline{topological space} $(M,\mathcal{O})$ by choosing a collection $\mathcal{O}$ of  (then-to-be-called \underline{open}) subsets of $M$, provided the choice has been made such that
\begin{enumerate}
\item the trivial subsets are open: $\emptyset, M \in \mathcal{O}$\,,
\item finite intersections of open sets are open: $U, V \in \mathcal{O}$ implies $U\cap V \in \mathcal{O}$\,,
\item arbitrary unions of open sets are open: $U_\alpha \in \mathcal{O}$ for all $\alpha\in A$ implies $\bigcup_{\alpha\in A} U_\alpha \in \mathcal{O}$\,.  
\end{enumerate}

\sep{\bf Remarks.}\lift
\begin{enumerate}
\item The choice of topology for any given set is far from unique. For sets of cardinality from one to seven, there are 1, 4, 29, 355, 6942, 209527 and 9535241 topologies, respectively, and this number increases rapidly with growing cardinality.
\item The coarsest topology any set $M$ can be equipped with is $\mathcal{O} = \{\emptyset, M\}$ (there is obviously no topology with fewer open sets), while the finest topology is $\mathcal{O}=\mathcal{P}(M)\equiv\{U \,|\, U \subset M\}$ (there is obviously no topology with more open sets). Useful choices for topologies usually lie between these extremes.
\item Only for finite sets $M$ may one provide a topology $\mathcal{O}$ by writing down the complete list of sets one chooses to be open. For inifinite sets one needs to resort to an indirect definition of the open sets and then prove that their collection $\mathcal{O}$ indeed makes $M$ into a topological space. This is the case for the important example that follows.
\end{enumerate}

\sep{\bf Definition.} A subset $U$ of $\mathbb{R}^d$ is called open in the \underline{standard topology} $\mathcal{O}_s$ on $\mathbb{R}^d$ if for every $x=(x^1, \dots, x^d)\in U$ there exists a positive real $\epsilon$ such that the ball
$$B_\epsilon(x) = \{(y^1, \dots, y^d) \in \mathbb{R}^d \,|\, \sum_{i=1}^d (x^i-y^i)^2 < \epsilon\}$$
lies entirely within $U$. 

\sep{\bf Remark. } We will tacitly assume in the following that $\mathbb{R}^d$ is equipped with the standard topology, unless explicitly stated otherwise.

\sep{\bf Exercise.} Prove that $(\mathbb{R}^d, \mathcal{O}_s)$ is a topological space.  

\sep{\bf Definition.} A map $\phi: M \to N$ between two topological spaces $(M,\O_M)$ and $(N,\O_N)$ is called \underline{continuous} if $\textrm{preim}_\phi(V) \equiv \{x\in M \,|\, \phi(x)\in V\} \in \O_M$ for every $V\in \O_N$. A bijection $\phi$ is called a \underline{homeomorphism} if both $\phi$ and $\phi^{-1}$ are continuous. Two topological spaces between which there exists a homeomorphism are called \underline{homeomorphic}. 

\sep{\bf Remarks. }\lift
\begin{enumerate}
\item It is easy to see that any map $\phi: M \to N$ is continuous if $N$ is equipped with the chaotic topology or if $M$ is equipped with the discrete topology. This is an example of the earlier claim that useful topologies lie somewhere between these extremes.
\item The definition of continuity of a function $f: \mathbb{R}\to\mathbb{R}$ reduces to the elementary $\epsilon$-$\delta$-criterion from undergraduate analysis.
\end{enumerate}

\sep{\bf Definition.} A topological space $(M,\O)$ is called a \underline{$d$-dimensional} \underline{topological manifold} if for every $p\in M$ there exists an open set $U\ni p$ and a homeomorphism $\phi: U \to \phi(U) \subset \mathbb{R}^d$. The component functions $\phi^1, \dots, \phi^d: U \to \mathbb{R}$ are then called the \underline{coordinate functions} on $U$, and for every point $q \in U$, the $d$-tuple $(\phi^1(q), \dots, \phi^d(q))$ is called the \underline{coordinates} of $q$, with respect to the \underline{chart} $(U,\phi)$. 

\sep{\bf Remarks.}\lift
\begin{enumerate}
   \item Covering a manifold by charts is what sailors do. This allows to study objects on the manifold in terms of their chart representatives. For instance, a curve $\gamma:  \mathbb{R} \to M$ (the course of the ship) on the manifold (in the real world) is represented in a chart $(U,\phi)$ by the curve $\gamma_\phi: \textrm{preim}_\gamma(U) \to \phi(U)$ with $\gamma_\phi = \phi \circ \gamma$ (the course drawn on the paper chart). 
 \item In general a single chart does not suffice to cover the entire manifold, as is well-known for maps of the Earth. In general one needs several overlapping charts, and one needs to know how to make transitions between these charts.
\end{enumerate}

\sep{\bf Definition.} Two charts $(U,\phi)$ and $(V,\psi)$ of a topological manifold are called \underline{$C^k$-compatible} if either (a) $U \cap V = \emptyset$ or (b) $U \cap V \neq \emptyset$ and the chart transition map
$$\psi \circ \phi^{-1}: \phi(U\cap V) \to \psi(U\cap V),$$
which by construction is map between open subsets on $\mathbb{R}^d$, is invertible, and both the transition map and its inverse are $k$-times continuously differentiable.
 A \underline{$C^k$-atlas} is a family of mutually $C^k$-compatible charts $(U_\alpha,\phi_\alpha)_{\alpha\in A}$ with $M = \bigcup_{\alpha\in A} U_\alpha$. A $C^k$-atlas is called \underline{maximal} if any chart $(N,\nu)$ that is $C^k$-compatible with all charts in the atlas is already contained in the atlas. A topological manifold $(M, \O)$ equipped with a maximal $C^k$-atlas is called a \underline{$C^k$-(differentiable) manifold}. 

\sep{\bf Remarks.}\lift
\begin{enumerate}
  \item Starting from a maximal $C^0$-atlas, charts need to be removed in order to obtain a maximal $C^1$-atlas, and even more charts need to be removed to obtain a $C^2$-atlas, and so forth. In practice, however, one simply specifies a maximal $C^k$-atlas $\mathcal{A}$ by specifying some (rather minimal than maximal) $C^k$-atlas $\mathcal{A}_0$, and then declaring any further chart that is $C^k$-compatible with every chart in $\mathcal{A}_0$ an element of $\mathcal{A}$.  
  \item On a $C^k$-manifold $(M,\O,\mathcal{A})$ one may now define a curve $\gamma: \mathbb{R} \to M$ to be \underline{$m$-times continuously differentiable, or $C^k$,} at parameter value $t_0\in \mathbb{R}$, if for some chart $(U,\phi)$ with $\gamma(t_0)\in U$ the chart representative $\gamma_\phi$ is $m$-times continuously differentiable in $t_0$ as a curve on $\mathbb{R}^d$. It is then precisely the mutual $C^k$-compatibility of all charts in the atlas $\mathcal{A}$ that guarantees that also the chart representative $\gamma_\psi$ of the curve with respect to any any other chart $(V,\psi)$ with $\gamma(t_0)\in V$ is $m$-times continuously differentiable at $t_0$, since
$$\gamma_\psi = \psi \circ \gamma = \psi \circ \phi^{-1} \circ \phi \circ \gamma = (\psi\circ \phi^{-1}) \circ \gamma_\phi$$ and the chart transition map $\psi\circ \phi^{-1}$ is $k$-times continuously differentiable.
Hence the above definition of $m$-times differentiable curves is independent of the actual chart representative, and thus well-defined.   
 \item It is easy to see that any two charts on a topological manifold are automatically $C^0$-compatible. Thus every topological manifold is a $C^0$-manifold, and vice versa. In particular, it depends on one's taste of whether one wishes to decide the continuity of a curve, say, directly at the level of the topological manifold in terms of its topology or, alternatively, at the level of charts in terms of the continuity of some chart representative. In contrast, $C^k$-differentiability for $k\geq 0$ can no more be decided at the topological manifold level, but one must descend to the level of charts. 
\end{enumerate}

\sep{\bf Exercise.} (a) Provide a definition of an $m$-times continuously differentiable function $f: M \to \mathbb{R}$ on a $C^{k\geq m}$-manifold and show that the definition is independent of the choice of chart. (b) Show that the set $C^k(M)$ of $k$-times continuously differentiable functions is made into a real vector space,  where addition of two functions and multiplication of a real number with a function are defined pointwise.   

\sep{\bf Example.} We illustrate the abstract theory developed here for the example of the real plane $M=\mathbb{R}^2$, which we make into a topological space by equipping it with the standard topology. But then it is already a topological manifold of dimension $2$ that can be covered by a single chart $(U,\phi)$ with $U=\mathbb{R}^2$ and $\phi: U \to \mathbb{R}^2$ defined by $\phi^1(x,y)=x$ and $\phi^2(x,y)=y$, which is of course continuous in the standard topology. Thus we have an atlas $\mathcal{A}$ consisting of only the chart $(U,\phi)$, which is trivially $C^\infty$-compatible with itself, so that $\mathcal{A}$ is a $C^\infty$-atlas. Extending $\mathcal{A}$ to a maximal atlas by assuming that any other chart $(V,\psi)$ that is $C^\infty$-compatible to $(U,\phi)$ is also contained, we have thus constructed a $C^\infty$-manifold $(\mathbb{R}^2,\mathcal{O},\mathcal{A}_\textrm{max})$. An interesting example of another chart in $\mathcal{A}_\textrm{max}$ is $V=\mathbb{R}^2\backslash\{(s,0)\,|\, s\in [0,\infty)\}$ and $\psi: V \to \mathbb{R}^2$ defined by $\psi^1(x,y)=\sqrt{x^2+y^2}$ and $\psi^2(x,y)=\arctan(y/x)$. The chart transition map is then 
$$\phi\circ \psi^{-1}: \mathbb{R}^+\times(0,2\pi) \to \mathbb{R}^2\backslash\{(0,s) \,|\, s\in[0,\infty) \}\,,\qquad
(\phi\circ\psi^{-1})(r,\varphi) = (r \cos \varphi, r \sin \varphi)\,,$$
which is clearly invertible and $C^\infty$. Hence indeed, $(V,\psi)$ is contained in the maximal atlas we constructed.

\newpage
\section*{Lecture II:\, TENSORS}
Without additional structure on a smooth manifold, the most general type of data one can define on a differentiable manifold are tensor fields, whose definition and properties we concisely introduce in this second lecture. Indeed, it will be tensor fields that will serve in the next lecture as the mathematical objects encoding prototypical  matter and the fundamental geometry on a smooth manifold. 

\sep{\bf Definition.} Let $M$ be a $C^\infty$-manifold and $\gamma$ a $C^1$-curve on $M$. Then the tangent vector $\dot\gamma_p$ to the curve $\gamma$ at the point $p=\gamma(t_p)$ is the linear map
$$\dot\gamma_p: C^{1}(M) \to \mathbb{R}, \qquad \dot\gamma_p f = \frac{d(f\circ \gamma)}{dt}(t_p)\,.$$

\sep{\bf Remarks.} 
\begin{enumerate}
\item The \underline{sum of two tangent vectors} at the same point $p$, defined by its action on an arbitrary $C^1$-function $f$ as
$$(\dot\gamma_p \oplus \dot \delta_p)f = \dot\gamma_p f + \dot\delta_p f\,,$$
is again a tangent vector, i.e., $\dot\sigma_p =\dot\gamma_p\oplus\dot\delta_p$ for some curve $\sigma: \mathbb{R} \to M$ through $p$. Such a curve is readily constructed using some chart $(U,\phi)$ with $p\in U$; for
$$\sigma(t) := \phi^{-1}\left((\phi\circ\gamma)(t) + (\phi\circ\delta)(t) - (\phi\circ\gamma)(t_p)\right)$$
one has $\sigma(t_p)=p$ and for any $C^1$-function $f$ 
$$\dot\sigma_p f = \partial_a(f\circ\phi^{-1})(\phi(p))\left(\frac{d \gamma_\phi{}^a}{dt}(t_p) + \frac{d \delta_\phi{}^a}{dt}(t_p)\right) = \dot\gamma_p f+\dot\delta_p f
$$
according to the multi-dimensional chain rule (and summing over repeated indices) for the function $f\circ\phi^{-1}$ and curves $\phi\circ\delta$ and $\phi\circ\delta$ on $\mathbb{R}^d$. Similarly, one defines the \underline{$S$-multiplication} $\odot$ of real number with a tangent vector and shows that the result is again a tangent vector.
\item The set $T_pM$ of all tangent vectors through a point $p$, equipped with $\oplus$ and $\odot$ constitutes a real vector vector space, the \underline{tangent space to $M$ at $p$}. 
\item A chart $(U,\phi)$ induces at each point $p\in M$ a particular basis of the tangent vector space $T_pM$, for then 
$$\dot\gamma_p f = \frac{d(f\circ \phi^{-1} \circ \phi \circ \gamma)}{dt}(t_p) = \frac{d(\phi\circ\gamma)^a}{dt}(t_p) \partial_a(f\circ\phi^{-1})(\phi(p))\,,$$
and defining the chart-dependent derivative operators $\frac{\partial}{\partial \phi^a}|_p: C^{1}(M) \to \mathbb{R}$ by their action 
$$\left.\frac{\partial}{\partial\phi_a}\right|_p f = \partial_a(f \circ \phi^{-1})(\phi(p))$$
on any $C^1$-function $f$ on $M$, we find that 
$$\dot\gamma_p = \frac{d \gamma_\phi^a}{dt}(t_p) \left.\frac{\partial}{\partial\phi^a}\right|_p\,.$$
Thus we see that a chart $(U,\phi)$ gives rise to the \underline{chart-induced basis} 
$$\D{\phi}{1}, \dots, \D{\phi}{d}$$ 
in each tangent space $T_pM$ if $p\in U$, and the \underline{components of the tangent vector} with respect to this basis are given by the derivative of the chart-representative $\gamma_\phi$ with respect to its curve parameter.
\item Under a change of chart from $(U,\phi)$ to $(V,\psi)$, the chart-induced basis vectors change by the linear transformation 
$$\D{\psi}{a} = (C^{\phi}_{\psi}|_p)^m{}_a \D{\phi}{m}\,,$$
where $(C^{\phi}_{\psi}|_p)^m{}_a = \partial_a(\phi\circ\psi^{-1})^m|_{\psi(p)}$. Accordingly, the components of a tangent vector in the new basis are
$$\frac{d\gamma_\psi{}^a}{dt}(t_p) = (C^{\psi}_{\phi}|_p)^a{}_m\frac{d\gamma_\phi{}^m}{dt}(t_p)\,,$$
where it should be noted that it now reads $C^\psi_\phi|_p$ instead of $C^\phi_\psi|_p$, and that these two matrices are related by inversion, $(C^\phi_\psi|_p)^m{}_a (C^\psi_\phi|_p)^a{}_n =\delta^m_n$.
\end{enumerate}

\sep{\bf Definition.} The \underline{differential} $d_pf$ of a $C^1$-function $f$ on $M$ is the linear map
$$d_pf: T_pM \to \mathbb{R}, \quad d_pf(X) = Xf\,.$$

\newpage
\sep{\bf Remarks}
\begin{enumerate}
  \item Given a chart $(U,\phi)$, the components $\phi^a: U \to \mathbb{R}$ with $a=1,\dots,d$ of the chart map are $C^1$-functions, and thus the differentials $d_p\phi^1, \dots, d_p\phi^d$ are $d$ elements of the so-called \underline{cotangent space} $T^*_pM$ at $p$, the dual vector space to $T_pM$. 
\item The action of the differentials of the components of the chart map act on  the chart-induced basis of $T_pM$ as
$$d_p\phi^a(\D{\phi}{b}) = \D{\phi}{b}\phi^a = \partial_b(\phi^a\circ \phi^{-1})(\phi(p)) = \delta^a_b\,.$$
Thus we find that the $d_p\phi^1, \dots, d_p\phi^d$ constitute a basis of $T^*_pM$, namely the dual basis to the chart-induced basis on the tangent space $T_pM$.  
\item Expanding a vector $v\in T_pM$ and a co-vector $\omega\in T_p^*M$ in the basis and dual basis induced by a chart $(U,\phi)$ containing $p$, 
$$v = v_\phi{}^b \D{\phi}{b}\qquad\textrm{and}\qquad \omega = \omega^\phi{}_a d_p\phi^a\,,$$
one finds that in terms of components, 
$$\omega(X) = \omega^\phi{}_a d_p\phi^a(v_\phi{}^b \D{\phi}{b}) = \omega^\phi{}_a v_\phi{}^b \delta^a_b = \omega^\phi{}_a v_\phi{}^a\,.$$ 
\item Under change of chart from $(U,\phi)$ to $(V,\psi)$, with both containing the point $p$, the differentials transform as
$$ d_p\psi^a = (C^\psi_\phi|_p)^a{}_b d_p\phi^b$$
\end{enumerate} 

\sep{\bf Definition.} 
On a $C^1$-manifold $M$, a \underline{tensor of valence $(r,s)$} at a point $p\in M$ is an element $t$ of the vector space
$$(T_pM)^r_s = \underbrace{T^*_pM \otimes \dots \otimes T^*_pM}_{r \textrm{ times}} \otimes \underbrace{T_pM \otimes \dots \otimes T_pM}_{s \textrm{ times}}\,.$$
The \underline{components} of $t$ with respect to a chart $(U,\phi)$ containing the point $p$ are the real numbers
$$t_\phi{}^{a_1 \dots a_r}{}_{b_1 \dots b_s} = t\left(d_p\phi^{a_1}, \dots, d_p\phi^{a_r},\D{\phi}{b_1}, \dots, \D{\phi}{b_s}\right)\,,$$
and the tensor can be reconstructed from its components by virtue of
$$t = t_\phi{}^{a_1\dots a_r}{}_{b_1\dots b_s} \D{\phi}{a_1} \otimes\dots\otimes \D{\phi}{a_r} \otimes d_p\phi^{b_1} \otimes \dots \otimes d_p\phi^{b_s}\,.$$

\newpage
\sep{\bf Remarks.} 
\begin{enumerate}
\item Recall that for finite-dimensional real vector spaces $V$ and $W$, and $v\in V$ and $w\in W$, the \underline{tensor product (of vectors)} $v\otimes w$ is the bilinear map
$$v \otimes w: V^* \times W^* \to \mathbb{R}, \quad (v\otimes w)(\nu,\mu) = v(\nu) w(\mu)\,,$$
and the \underline{tensor product (space)} is the real vector space
$$V\otimes W = \{ a^{ij} v_i\otimes w_j \,|\, i=1,\dots,\dim V;\,j=1,\dots,\dim W\}$$
(where $v_1,\dots,v_{\dim V}$ and $w_1,\dots,w_{\dim W}$ are bases of $V$ and $W$, respectively), equipped with the addition of bilinear maps and their multiplication with real numbers.
\item According to the above definition, $(T_pM)^r_s$ equipped with the addition of multi-linear maps and their multiplication with real numbers is a vector space of dimension $(\dim M)^{r+s}$. Further, a co-vector $\omega\in T^*_pM$ is a tensor of valence $(0,1)$. By virtue of $V=(V^*)^*$ for any finite-dimensional vector space $V$, one also sees that a vector $X\in T_pM$ is a tensor of valence $(1,0)$. It is customary to say that a scalar (real number) is a tensor of valence $(0,0)$ at $p$.  
\item Under a change of chart from $(U,\phi)$ to $(V,\psi)$, the components of a tensor of valence $(r,s)$ change as
$$t_\psi{}^{a_1 \dots a_r}{}_{b_1\dots b_s} = (C^\psi_\phi)^{a_1}{}_{m_1} \dots (C^\psi_\phi)^{a_r}{}_{m_r} (C^\phi_\psi)^{n_1}{}_{b_1} \dots (C^\phi_\psi)^{n_s}{}_{b_s} t_\psi{}^{m_1 \dots m_r}{}_{n_1\dots n_s}\,.$$
\end{enumerate}

\sep{\bf Definition.} The \underline{tensor bundle} $(TM)^r_s$ over a manifold $M$ is the set
$$(TM)^r_s = \{(p,t) \,|\, p \in M \textrm{ and } t \in (T_pM)^r_s\}$$
together with the canonical projection map
$\pi: TM \to M$ defined by $\pi(p,t) = p$. A \underline{tensor field} $T$ is a map $T: M \to (TM)^r_s$ for which $(\pi \circ T)(p) = p$ for all points $p \in M$.

\sep{\bf Remarks.} 
\begin{enumerate}
  \item A chart $(U,\phi)$ induces on its domain the $(1,0)$-tensor (vector) fields
$$\DD{\phi}{a}: U \to (TU)^1_0\,, \qquad p \mapsto \left(p,\D{\phi}{a}\right)\,,$$
and the $(0,1)$-tensor fields (co-vector)
$$d\phi^a: U \to (TU)^0_1\,, \qquad p \mapsto \left(p,d_p\phi^a\right)\,.$$
\item Expressing a tensor field $T$ is terms of its \underline{component functions} with respect to the above tangent and cotangent basis fields, 
$$T_\phi(p) = T_\phi{}^{a_1\dots a_r}{}_{b_1\dots b_s}(\phi(p)) \DD{\phi}{a_1} \otimes\dots\otimes \DD{\phi}{a_r} \otimes d\phi^{b_1} \otimes \dots \otimes d\phi^{b_s}\,,$$
$T$ is called a \underline{$C^m$-tensor field} if the above tensor component functions $T^{a_1 \dots a_r}{}_{b_1 \dots b_s}: \phi(U) \to \mathbb{R}$ are $m$-times continuously differentiable. 
\end{enumerate} 


\newpageja
\section*{Lecture III: \, HYPERBOLIC GEOMETRIES}
In this lecture, we start our investigation of which geometries on a smooth manifold can serve as a geometry of spacetime. We will start from the most general assumption of the geometry being given by some arbitrary tensor field, and identify crucial properties such a tensor field must satisfy in order to present a viable spacetime geometry. This will culminate in our final definition of a spacetime in lecture V. 

\sep{\bf Definition.} A \underline{geometry} on a smooth manifold $M$ is given by a smooth tensor field $G$, which may be restricted by further conditions. 

\sep{\bf Examples.} 
\begin{enumerate}
  \item The traditionally most intensively studied geometry is \underline{Riemannian geometry}, given by a $(0,2)$-tensor field $g$ that is restricted to be symmetric ($g(X,Y)=g(Y,X)$ for all vectors $X,Y$ at each point) and positive definite ($g(X,X)>0$ for every non-vanishing vector $X$ at each point).  
The spacetime geometry in general (and hence special) relativity is a \underline{Lorentzian} \underline{geometry}, given by a $(0,2)$-tensor $g$ that is symmetric and of signature $(1,\dim M -1)$. 
\item The geometry of the phase space of classical mechanical systems carries a \underline{symplectic} \underline{geometry}, given by a $(0,2)$-tensor field $\omega$ that is restricted to be anti-symmetric ($\omega(X,Y)=-\omega(Y,X)$ for all vectors $X, Y$ at each point), non-degenerate ($\omega(X,Y)$ for all vectors $X$ at a point already implies that $Y=0$.) and closed ($\DD{\phi}{a} \omega_{bc} + \DD{\phi}{b} \omega_{ca} + \DD{\phi}{c} \omega_{ab} = 0$).  
\item As one example for a spacetime geometry beyond Lorentzian manifolds, we will meet \underline{area metric geometry}, given by a $(0,4)$-tensor field $G$ that features the symmetries
$$G(X,Y,A,B)=G(A,B,X,Y) \qquad \textrm{and} \qquad G(A,B,X,Y)=-G(B,A,X,Y)$$
and is non-degenerate in the sense that $G(A,B,X,Y) = 0$ for all linearly independent pairs of vectors $A,B$ already implies that the pair of vectors $X, Y$ is linearly dependent. Employing the techniques we will develop in these lectures, we will see that in order to serve as a physically viable spacetime structure, an area metric needs to be restricted by further algebraic conditions (which are precisely the same deep conditions that lead to the Lorentzian signature condition in the metric case).   
\end{enumerate}  

\sep{\bf Remarks.}
\begin{enumerate}
  \item A geometry is called \underline{flat} if there exists an atlas such that in each chart the component functions of the tensor field $G$ are constant. Unless explicitly stated, however, in these lectures we will \underline{not} make the assumption that the geometry $(M,G)$ is flat. 
\item In Riemannian geometry $(M,g)$, a necessary and sufficient criterion for flatness is that the Riemann-Christoffel tensor associated with the metric $g$ vanishes. For a symplectic geometry $(M,\omega)$, in contrast, it is a fundamental result that one can always find ``Darboux'' charts, where the components of the symplectic form $\omega$ are constants, so that every symplectic manifold is flat.
\end{enumerate}

\sep{\bf Definition.} \underline{Probing matter} on a smooth manifold $M$ equipped with a geometry $G$ is given by a tensor field $\Phi$ that takes its values in some 
vector subspace $V$ of a $(TM)^r_s$ and satisfies a linear partial differential equation
$$\left[\sum_{n=1}^s Q_{MN}^{i_1 \dots i_n}(x) \, \partial_{i_1} \dots \partial_{i_n}\right] \Phi^N(x)= 0\,,$$
where the coefficients $Q$ are constructed solely from the components of $G$ and its partial derivatives such that (a) the entire equation transforms as a tensor and (b) the initial-value problem is well-posed. Here the small Latin indices run over $1, \dots, \dim M$ and the capital Latin indices run over $1, \dots, \dim V$.

\newpage
\sep{\bf Remarks.} 
\begin{enumerate}
  \item The terminology ``probing matter'' refers to the linearity of the differential equation. For we will see that studying the properties of a tensor field governed by a linear differential equation, one can learn important lessons about the underlying geometry. 
While at first sight one could learn similar lessons starting from non-linear equations, it would often be impossible to disentangle the properties of the matter field from properties of the underlying geometry. 
  \item A restriction of $\Phi$ to a proper vector subspace $V$ is effected by linear conditions on $\Phi$, such as symmetry conditions. For instance, while a generic $(0,2)$-tensor $\Phi$ takes its values in a $(\dim M)^2$-dimensional tensor space, a symmetric one takes its values in a subspace $V$ with $\dim V = \dim M(\dim M + 1)/2$.
\item Under a change of chart, only the highest order coefficient $Q_{MN}^{i_1 \dots i_s}$ transforms as a tensor with respect to all its small and capital Latin indices. The lower order coefficients however generically pick up additional terms involving higher order coefficients, and thus do not transform as tensors. 

\item The \underline{initial-value} problem is well-posed if there exist hypersurfaces such that the prescription of initial data on these surfaces uniquely determines, by virtue of the above differential equation, the value of $\Phi$ at every point of the manifold. This prediction of the future, or post-diction of the past, is the essence of classical physics.
\end{enumerate}

\sep{\bf Theorem.} If the linear partial differential equations of the previous definition have a well-posed initial value problem, then
$$P(x,k) = \pm \rho \det_{M,N}\left[Q^{i_1\dots i_s}_{MN}(x) k_{i_1} \dots k_{i_s}\right]$$ 
defines a positive hyperbolic homogeneous polynomial in each cotangent space $T^*_xM$, where $\rho$ is a scalar density constructed from the geometric tensor $G$ such that $P(x,k)$ is a co-tangent bundle function and 
\begin{enumerate}
  \item \underline{homogeneous} means that for any $\lambda$ and any co-vector $k\in T^*_xM$ one has $P(x,\lambda k)=\lambda^{\textrm{deg } P} P(x,k)$, where $\deg P$ is called the \underline{degree} of P
  \item \underline{hyperbolic} means that there exists at least one co-vector $h$ (called a hyperbolic co-vector of $P$ at $x\in M$) with $P(x,h) \neq 0$ such that for every co-vector $q \in T^*_xM$ the polynomial equation
$$P(x,q + \lambda h) = 0$$
in the real number $\lambda$ has $\deg P$ many (and thus only) real solutions. 
\item \underline{positive} means that the sign $\pm$ is chosen such that $P(h)>0$.
\end{enumerate} 

\begin{figure}[h]
  \includegraphics[width=10cm]{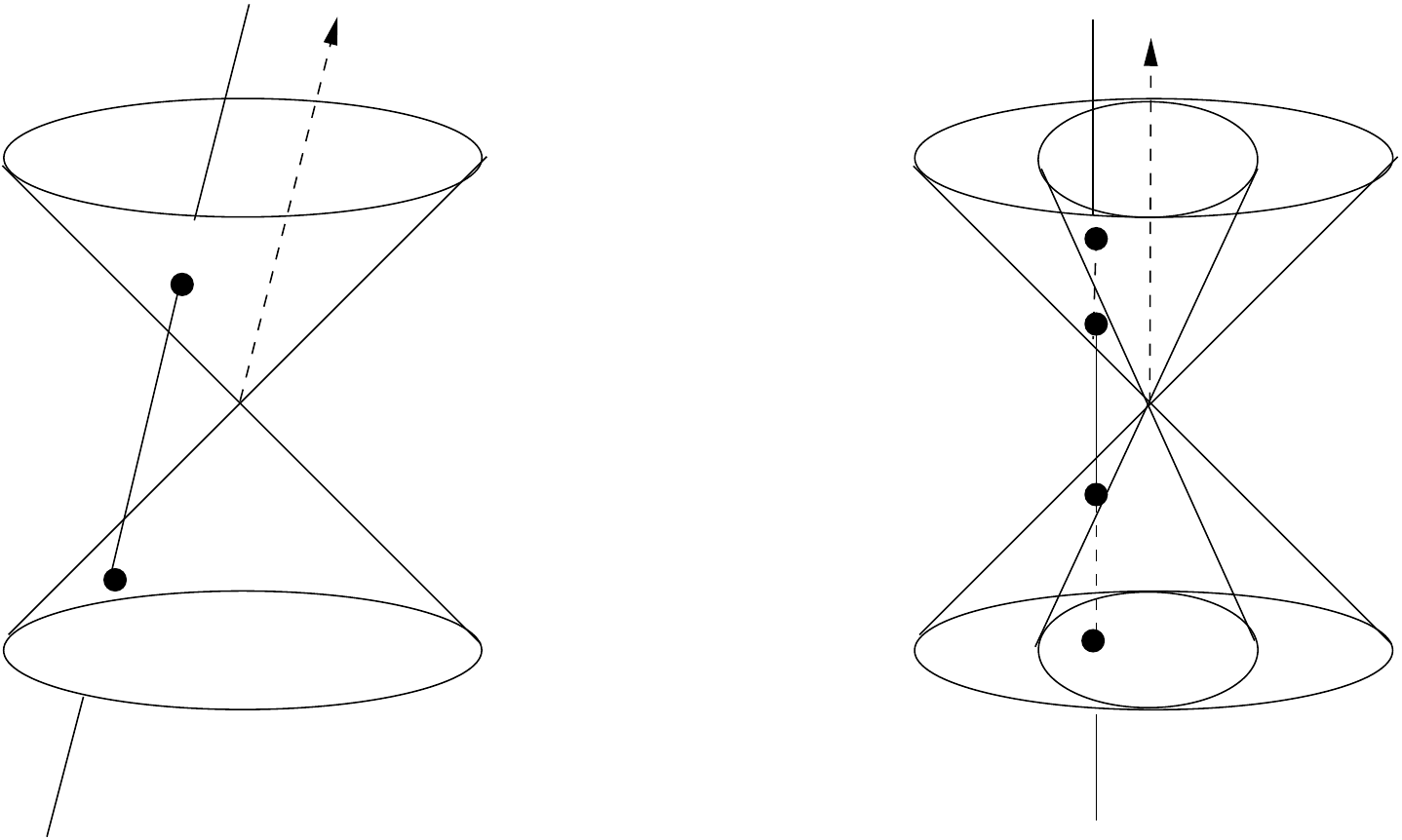}
  \caption{A co-vector is hyperbolic if any line in its direction intersects the vanishing set of the homogeneous polynomial $P$ exactly $\deg P$ many times; shown examples are for degree two and four, respectively.}
\end{figure}

\sep{\bf Remarks.} 
\begin{enumerate}
  \item The scaling of the co-tangent bundle function introduced by some choice of scalar density $\rho$ does not affect the hyperbolicity condition, or in fact any of the other conditions we will reveal in the course of these lectures to be imposed on a geometry to define a bona fide spacetime structure.   
 \item Only a hypersurface $\Sigma$ whose \underline{co-normals} at each point $y\in\Sigma$ (i.e., co-vectors $n\in T^*_yM$ such that for every smooth curve $\gamma$ on $\Sigma$ through $y$, $n(\dot\gamma_y)=0$) are hyperbolic co-vectors can serve as initial data surfaces for the underlying field equations. The condition that the polynomial be hyperbolic thus amounts to the condition that there be any viable initial data surfaces.
\item The requirement that $P$ be hyperbolic clearly imposes a restriction on the underlying geometry $G$. One main topic of these lectures is to study the full extent of this restriction, and to employ these insights in order to identify all geometries that can serve as a classical spacetime structure.  
\end{enumerate}

\sep{\bf Example.}
For the Klein-Gordon equation on a Lorentzian manifold $(M,g)$,
$$g^{ab}(x)\partial_a\partial_b \Phi(x) - \frac{1}{2} g^{ab}(x) g^{ms}(x)(\partial_a g_{sb}(x) + \partial_b g_{as}(x) - \partial_s g_{ab}(x)) \partial_m \Phi(x) + m^2 \Phi(x) = 0\,,$$
 one finds $P(x,k)=g^{ab}(x) k_a k_b$, which is hyperbolic if and only if $g$ is of Lorentzian signature $(1,d-1)$ or $(d-1,1)$. The positivity requirement then narrows this down to the signature $(1,d-1)$. This means that the Klein-Gordon equation cannot have a well-posed initial value problem unless formulated on a Lorentzian geometry. The hypersurfaces with hyperbolic co-normal are then precisely the so-called spacelike hypersurfaces, and indeed it is a well-known fact that every initial data surface for the Klein-Gordon equation is spacelike (but the converse does not hold). The previous theorem generalizes this to any tensorial geometry.

\begin{figure}[h]
  \includegraphics[width=10cm]{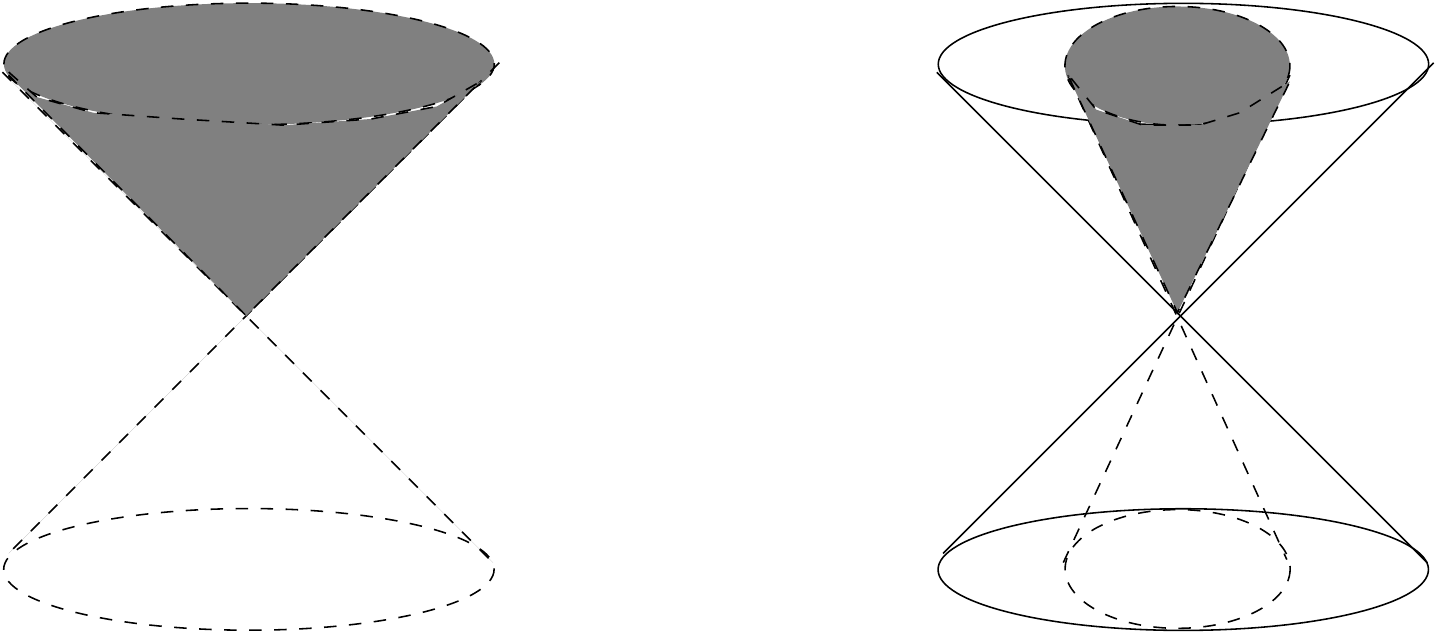}
  \caption{Hyperbolicity cones $C(P,h)$ for the previously shown examples of degrees two and four, respectively.}
\end{figure}

\sep{\bf Definition.} If $h$ is a hyperbolic co-vector of $P$, then there is an entire connected set of hyperbolic co-vectors, the so-called \underline{hyperbolicity cone} $C(P,h)$ around $h$. The hyperbolicity cone is explicitly obtained by the following construction. From the coefficients $h_0, \dots, h_{\textrm{deg } P}$ of the expansion 
$$P(x,q + \lambda h) = h_0(x,q,h)\, \lambda^{\deg P} + h_1(x,q,h)\, \lambda^{\deg P - 1} + \dots + h_{\deg P}(x,q,h)$$
one constructs the matrices $H_1, H_2, \dots, H_{\deg P}$ as
\begin{equation}
  H_i(q,h) = \left[\begin{array}{ccccc}
    h_1 & h_3 & h_5 & \dots & h_{2i-1} \\
    h_0 & h_2 & h_4 & \dots & h_{2i-2} \\
    0   & h_1 & h_3 & \dots & h_{2i-3} \\
    0   & h_0 & h_2 & \dots & h_{2i-4} \\
    \vdots  &  \vdots  &  \vdots  &  \vdots  &  \vdots\\
    0 & 0 & 0 & \dots & h_i 
\end{array}\right]_{i \times i} \qquad \textrm{where } h_j \textrm{ is set to } 0 \textrm{ for } j>i\,.\nonumber
\end{equation}
Then the hyperbolicity cone around $h$ is the set
\begin{equation}
  C(P,h) = \{ h' \in T^*_xM \, |\,   \det H_i(h',h) > 0 \textrm{ for all } i=1,\dots,\deg P\}\,.\nonumber
\end{equation}

\sep{\bf Remarks.}
\begin{enumerate}
\item From its definition, it is clear that every $h'\in C(P,h)$ represents the same hyperbolicity cone as $h$ does, $C(P,h')=C(P,h)$, and that any hyperbolicity cone is \underline{open} in the standard topology on cotangent space. Garding further showed that $C(P,h)$ is indeed a \underline{convex cone} since for every positive real 
$$\lambda h' \in C(P,h) \qquad \textrm{ and } \qquad h' + h'' \in C(P,h)\,,$$
positive real $\lambda$ and any $h', h'' \in C(P,h)$. 
\item While $P$ is, by definition, positive on the entire cone $C(P,h)$, it can be shown to vanish on its boundary $\partial C$. 
\item If $P$ factorizes into lower degree polynomials 
$$P(x,k) = P_1(x,k)^{\alpha_1} \dots P_f(x,k)^{\alpha_f}\,,$$
then $P$ is hyperbolic with respect to $h$ if and only if every $P_1, \dots, P_f$ is hyperbolic with respect to $h$. For technical reasons, we will assume in the following that unless $\alpha_1 = \dots = \alpha_f = 1$, we will replace $P$ at each point $x\in M$ by the \underline{reduced polynomial}
$$P(x,k) =  P_1(x,k) \dots P_f(x,k)\,.$$
In any case, for the hyperbolicity cones we have
$$C(P,h) = C(P_1,h) \cap \dots \cap C(P_f,h)\,.$$
\end{enumerate}

\sep{\bf Example.} For a hyperbolic polynomial $P(k)=g^{-1}(k,k)$ defined in terms of a Lorentzian metric of signature $(+-\dots-)$, one finds that any co-vector $h$ with $g^{-1}(h,h)>0$ is a hyperbolic co-vector. Picking one such $h$, the expansion
$$P(q+\lambda h) = \lambda^2 g^{-1}(h,h) + \lambda 2 g^{-1}(q,h) + g^{-1}(q,q)$$
identifies the quantities
$$h_0(q,h) = g^{-1}(h,h),\qquad h_1(q,h)= 2 g^{-1}(q,h),\qquad h_2(q,h) = g^{-1}(q,q)$$
and thus 
$$\det H_1(q,h) = \det\left[ 2 g^{-1}(q,h) \right] \quad\textrm{ and } \quad \det H_2(q,h) = \left[\begin{array}{cc} 2 g^{-1}(q,h) & 0 \\ g^{-1}(h,h) & g^{-1}(q,q)\end{array}\right]\,,$$
so that the hyperbolicity cone of $P$ that contains $h$ is the set
$$C(P,h) = \{h' \in T^*_xM \, | \, g^{-1}(h',h')>0 \, \textrm{ and } \, g^{-1}(q,h)>0\}\,.$$
\vspace{.1cm}


\newpageja
\section*{Lecture IV: \, MASSLESS DISPERSION}
In this lecture we reveal that the hyperbolic polynomial provides the dispersion relation for the underlying matter field equations in the high-energy (or,  equivalently, massless) limit. We employ this insight to derive the action for a massless point particle in terms of an associated dual polynomial, which in turn will play an important role in the definition of spacetimes.    

\sep{\bf Definition.} A family \underline{locally wavelike solutions} of linear matter field equations is a family of solutions of the form
$$\Phi^N_\lambda(x) = A^N_\lambda(x) e^{i S(x)/\lambda} \qquad \textrm{ for } \phi(x) \in U\,,$$
where $(U,\phi)$ is some conveniently small chart of the manifold, $\lambda$ is a positive real parameter, the \underline{phase function} $S$ is a smooth real function whose differential $dS$ is everywhere non-zero, and the real \underline{amplitude tensor field components} $A^N_\lambda$ admit Taylor expansions of the form
$$A^N_\lambda = \sum_{n=0}^\infty a^N_n(x) \lambda^n\,,\qquad \textrm{ where } \,\, a_0(x) \neq 0.$$  

\sep{\bf Remarks.}\lift
\begin{enumerate}
  \item A hypersurface along which $S=const.$ is called a \underline{wave front}, and the differential $d_pS/\lambda$ the \underline{wave co-vector} at $p\in U$. In the course of this lecture, we will learn how to associate a \underline{ray vector} with each \underline{wave co-vector} in suitable geometries.
  \item 
Insertion of the locally wavelike solution ansatz into the partial differential field equations yields 
$$Q_{MN}^{i_1 \dots i_s}(x) \partial_{i_1} \dots \partial_{i_s} S(x) \, a^N_0(x) + (\dots)_M \lambda + (\dots)_M \lambda^2 + \dots = 0\,,$$
which shows that for a locally wavelike solution one needs, to lowest order in $\lambda$, 
$$\det_{M,N}\left(Q_{MN}^{i_1 \dots i_s}(x) \partial_{i_1}S(x) \dots \partial_{i_s} S(x) \right) = 0\,.$$ 
In terms of the cotangent bundle function $P$ defined in the previous lecture, this means that to lowest order in $\lambda$, the wave co-vector $k=dS/\lambda$ of a locally wavelike solution must satisfy the \underline{dispersion relation}
\begin{equation}
P(x, k) = 0\,.\nonumber
\end{equation}
\item By taking into account more than just the lowest order, or even all orders in $\lambda$, the above dispersion relation may or may not be modified if one finally takes the high-frequency limit $\lambda\to 0$. If such a modification occurs in the high-frequency limit, we call the underlying field equations \underline{massive}, and otherwise \underline{massless}.
\end{enumerate} 

\sep{\bf Example.} We illustrate the derivation of the dispersion relation for a given field equation for the example of a Klein-Gordon equation in flat Lorentzian spacetime, which in a suitably chosen chart takes the simple form
$$\eta^ab \partial_a \partial_b \Phi(x) - m^2 \phi(x) = 0\,, \qquad \textrm{ where } m\geq 0\,.$$
Insertion of the locally wavelike solution ansatz yields 
$$-\eta^{ab} \partial_a S\partial_b S A_\lambda + \left[ 2i \eta^{ab} \partial_a S\partial_b A_\lambda + i \eta^{ab}\partial_a \partial_b  A_\lambda\right] \lambda + \left[\eta^{ab}\partial_a\partial_b A_\lambda + m^2 A_\lambda\right] \lambda^2 = 0 $$
so that $S(x) = k_a x^a$ solves the equation to lowest order in $\lambda$ for any co-vector $k$ with $\eta^{ab}k_a k_b = 0$, and the equation reduces to
$$\sum_{j=0}^\infty \left[\eta^{ab}\partial_a\partial_b a_{j-1} + m^2 a_{j-1} + 2 I \eta^{ab} k_a \partial_b a_j\right] \lambda^{j-1}\,,$$
where we now expanded the functions $A_\lambda$ in terms of the functions $a_j$ and defined $a_{-1}=0$. This is solved for arbitrary $\lambda$ by 
$$a_n(x) = a \frac{1}{n!} \left(\frac{i m^2 l_c x^c}{2 \eta^{ab}k_a l_b}\right)^n \qquad \textrm{ and thus } \qquad A_\lambda(x) = a e^{i \frac{\lambda m^2}{2 \eta^{ab} l_a k_b} l_c x^c}$$
for any real constant $a$ and co-vector $l$ such that $\eta^{ab} l_a l_b = 0$ but $\eta^{ab} l_a k_b \neq 0\,.$ In order to keep $A_\lambda$ real, one needs to absorb the phase into the phase function $S$ and thus obtains 
$$S(x) = \left(1 + \frac{\lambda^2 m^2}{2 \eta^{ab} l_a k_b}\right)k_c x^c \qquad \textrm{ and } \qquad A(x) = a\,.$$
But from this we obtain, having taken into account all orders of $\lambda$, the dispersion relation
$$P(x, \partial S(x)) = m^2\,.$$
Thus we identify the Klein-Gordon equation with $m=0$ as massless, and the  Klein-Gordon equation with $m>0$ as massive. 

\sep{\bf Definition.} The set of \underline{massless momenta} at a point $x\in M$ is the cone 
\begin{equation}
  N_x = \{k \in T_x^*M \,|\, P(x,k)=0 \}\,,\nonumber
\end{equation}

\begin{figure}[h]
  \includegraphics[width=10cm]{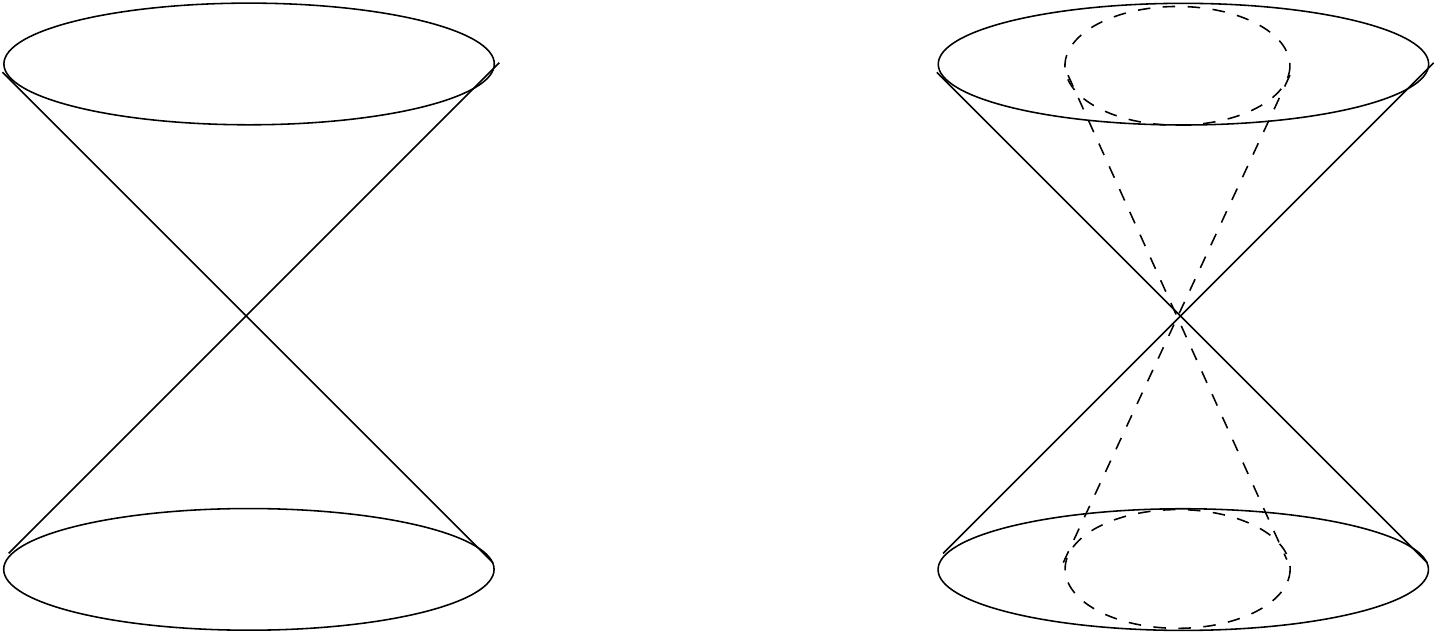}
  \caption{Massless momenta for a polynomial of degree two and four, respectively.}
\end{figure}

\sep{\bf Remarks.}\lift
\begin{enumerate}
\item For technical precision, we will occasionally focus on the \underline{smooth subcone}   
\begin{equation}
  N^\textrm{smooth}_x = \{ k \in  N_x \,|\, DP(x,k)\neq 0\}\,, \nonumber
\end{equation} 
where $DP$ denotes the derivative of the reduced $P$ with respect to the cotangent fibre.
\item While the massless momentum cone $N_x$ at each point is determined by the polynomial $P(x,\cdot)$, the converse question -- namely under which conditions the massless momentum cone $N_x$ at a point $x$ determines the polynomial $P_x$ up to a constant factor, is subtle, but of central importance. Indeed, it can be shown that for a cone $N_x$ of massless momenta, the polynomial is determined up to scale, since the real Nullstellensatz
$$\mathcal{I}(N_x) = \{ \alpha P_x \, |\, \alpha \in \mathbb{R}\}$$
holds for any reduced hyperbolic polynomial $P$. Here for any subset $S\subset T_x^*M$ of cotangent space, $\mathcal{I}(S)$ denotes the set of all polynomials on $T_x^*M$ that vanish on $S$. This real Nullstellensatz will become important shortly.
\end{enumerate}

In order to associate velocity vectors with massless particle momenta in physically meaningful fashion, we employ the dynamics of free massless point particles. 

\sep{\bf Theorem.} The action of a \underline{free} and \underline{massless} point particle is
\begin{equation}
  I_0[x,q,\lambda] = \int d\tau \left[q_a \dot x^a + \lambda P(x,q)\right]\,,\nonumber
\end{equation} 
where the function $\lambda$ is a Lagrange multiplier. 

\sep{\bf Remarks.}
\begin{enumerate}
  \item It is clear by construction that the above action describes a free and massless point particle. Note, in particular, that the geometry enters the action only through the dispersion relation enforced in the Lagrange multiplier term. 
  \item  In the following, we wish to eliminate the momentum $q$ and the Lagrange multiplier $\lambda$ to obtain an equivalent action in terms of the particle trajectory $x$ only. Variation of the Helmholtz action with respect to $\lambda$ of course enforces the null condition for the particle momentum. Now variation with respect to $q$ yields $\dot x = \lambda \, DP_x(q)$ for all $q\in N^\textrm{smooth}$,
 which implies the weaker equation
\begin{equation}
  [DP_x(q)] = [\frac{\dot x}{\lambda}]\,,\nonumber
\end{equation} 
where $[X]$ denotes the projective equivalence class of all vectors collinear with the vector $X$. It is this latter equation that we will need to to invert in order to eliminate the momentum from the above Helmholtz action. 
\end{enumerate}

\sep{\bf Definition.} The polynomial $P_x^\#: T_xM \to \mathbb{R}$ is called \underline{dual polynomial} to an \underline{irreducible} polynomial $P: T_x^* \to \mathbb{R}$ if
$$P_x^\#(DP_x(N^\textrm{smooth}_x))=0\,,$$
and is thus determined up to a real scale. 
For a polynomial $P_x$ that is \underline{reducible} into irreducible factors we define the dual polynomial as the product
\begin{equation}
(P_1(x,k)\cdots P_f(x,k))^\#(x,v) = P_1^\#(x,v) \cdots P_f^\#(x,v)\,,\nonumber
\end{equation}
of the duals $P^\#_i$ of the irreducible $P_i$, whence $P^\#$ is uniquely determined up to a real scale and also satisfies the previous equation.

\sep{\bf Remarks.} 
\begin{enumerate}
  \item The \underline{dual null cone} $N_x^\#$, defined as the image of the massless co-vector cone $N_x$ under the gradient map 
$$DP: N_x \to T_x^M\,, \qquad k \mapsto \frac{\partial P}{\partial k_a}(x,k)\,,$$
 is the vanishing set of $P_x^\#$. This is the geometric meaning of the dual polynomial.
  \item The \underline{existence} of a dual $P_x^\#$, and indeed its algorithmic computability for any reduced hyperbolic polynomial $P$, ultimately hinges on the real Nullstellensatz mentioned before. The the degree of $P^\#$, however, is generically different from the degree of $P$.
\item In principle, the \underline{construction} of the dual polynomial to a reduced hyperbolic polynomial may always be performed using Buchberger's algorithm. The bad news is that, in practice, such a direct calculation of dual polynomials of higher degree and in several variables (that is precisely the cases we are interest in) using elimination theory exhausts the capability of current computer algebra systems. The goods news, however, is that in some cases of physical interest one is nevertheless able to guess the dual polynomial by physical reasoning and then to readily verify it mathematically. In any case, since a dual polynomial always exists for the reduced hyperbolic polynomial we are considering here, we will simply assume in the following that a dual $P^\#$ has been found by some method.
\end{enumerate}

\sep{\bf Definition and Theorem.}
The \underline{Gauss map} 
$$[DP]: [N_x^{smooth}] \to [N^\#], \qquad [q] \mapsto [DP_x(q)]$$
and the \underline{dual Gauss map}
$$[DP^\#]: [N_x^{\# \,smooth}] \to [N_x], \qquad [X] \mapsto [DP_x^\#(X)]$$
are \underline{partial inverses} of each other, in the sense that for null co-vectors $k\in N_x^\textrm{smooth}$ 
\begin{equation}
 [DP_x^\#]([DP_x]([k])) = [k]  \qquad \textrm{ if }\, \det(DDP_x)(k)\neq 0\,,\nonumber
\end{equation}
and similarly with $P$ and $P^\#$ exchanged.

\sep{\sl Proof.} Writing the defining equation for the dual polynomial in the form
\begin{equation}
  P^\#(x,DP(x,k)) = Q(k) P(k) \qquad \textrm{ for all co-vectors } k\,,\nonumber
\end{equation}
(since this form does not require an explicit restriction to null co-vectors), differentiation with respect to $k$ yields, by application of the chain rule and then of Euler's theorem
$$DP(k) k=(\deg P) P(k)$$
for the homogeneous function $P$ on the right hand side, for any null co-vector $k$ satisfying the non-degeneracy condition $\det(DDP_x)(k)\neq 0$ that
\begin{equation}
  DP^\# (x,DP(x,k)) = \frac{Q(x,k)}{\deg P - 1} k  \,,\nonumber
\end{equation}
which in projective language is the statement of the theorem.

\sep{\bf Remarks.} \lift 
\begin{enumerate}
   \item We may thus solve, up to a real scale, the momentum-velocity relation $[DP_x(q)]=[\dot x/\lambda]$ for the massless particle for the momentum,
$$[q] = [DP_x^\#]([\dot x/\lambda])\,,$$
and obviously the homogeneity of $DP_x^\#$ in conjunction with the projection brackets \new{allows} to disregard the function $\lambda$ altogether. Translating this result back to non-projective language, another undetermined function $\mu$ appears,
\begin{equation}\label{nullmap}
  q = \mu\,  DP_x^\#(\dot x)\,.\nonumber
\end{equation}   
\item This reveals the physical meaning of the Gauss map $[DP_x]$ and its inverse $[DP_x^\#]$: up to some irrelevant conformal factor, they associate null particle momenta in $N_x^\textrm{smooth}$ with the associated null particle velocities in $N_x^{\#\textrm{ smooth}}$. 
\item Replacing the momentum in the action for the massless particle, and using again Euler's theorem, but now applied to the  homogeneous polynomial $P_x^\#$, one obtains the massless point particle action
\begin{equation}\label{nullaction}
  I_0[x,\mu] = \int d\tau \mu\, P^\#(x,\dot x)\,.\nonumber
\end{equation}
The \new{automatic} appearance of a final Lagrange multiplier $\mu$ also hardly comes as a surprise, since it is needed to enforce the null constraint $P_x^\#(\dot x)=0$.
This reveals the direct physical relevance of the dual tangent bundle function $P^\#$ as \new{} the tangent space geometry seen by massless particles. 
\end{enumerate}

\newpageja
\section*{Lecture V: \, SPACETIMES}
Finally we are in the position to write down a definite definition of what constitutes a spacetime geometry. The requirements of that definition are, on the one hand, necessary in order to ensure that the equations for the probing matter are predictive and that (as we will show) all observers agree on the distinction of particles and anti-particles; on the other hand, these requirements will be sufficient to develop the full kinematical apparatus known from general relativity, but for any spacetime geometry in the general sense discussed here.  

\sep{\bf Definition.} The polynomial $P_x: T_xM \to \mathbb{R}$ is called \underline{bi-hyperbolic} if both $P_x$ and its dual polynomial $P_x^\#$ are hyperbolic polynomials. 

\sep{\bf Remarks.} 
\begin{enumerate}
  \item Hyperbolicity of a polynomial on tangent space is defined precisely as hyperbolicity of a polynomial on cotangent space, but with all co-vectors replaced by vectors. 
  \item Hyperbolicity does not imply bi-hyperbolicity. However, Hyperbolic polynomials $P(k)=g^{-1}(k,k)$ of degree two (necessarily defined by an inverse Lorentzian metric $g^-1$) present an exception; their duals $P^\#(X) = g(X,X)$ are hyperbolic if and only if $P$ is hyperbolic. This is simply due to the obvious fact that a metric $g$ has Lorentzian signature if and only if its inverse $g^{-1}$ has Lorentzian signature. 
\end{enumerate}

\sep{\bf Definition.} Let $T$ be a nowhere vanishing smooth vector field on $M$ such that for each point $x\in M$ the vector $T(x)$ lies within one of the hyperbolicity cones of $P_x^\#$, which cone we then denote by $C^\#_x$. The thus defined smooth distribution of cones
$$C^\# = \{(x,C_x^\#) \, | \, x\in M\}$$
is then called a \underline{time orientation} of the manifold $(M,G)$.

\begin{figure}[h]
  \includegraphics[width=10cm]{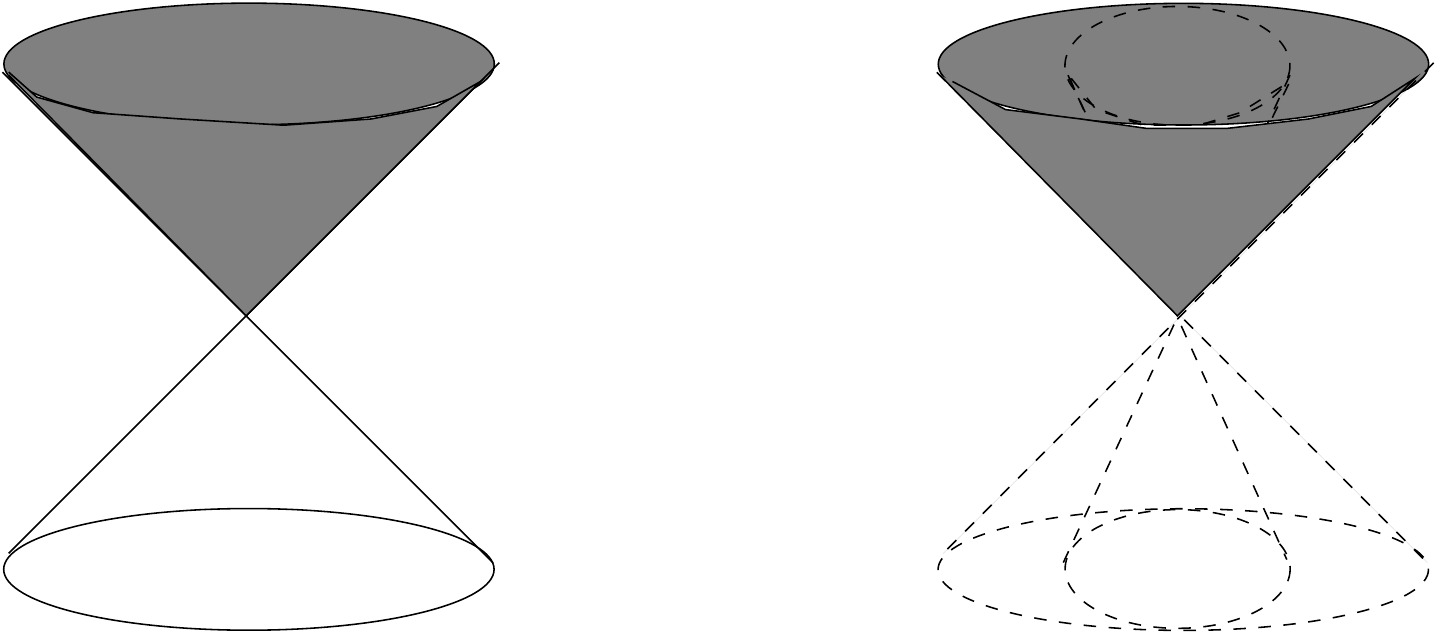}
  \caption{Time-orientation $C^\#$ in tangent space for polynomials of degree two and four, respectively.}
\end{figure}

\sep{\bf Remarks.}
\begin{enumerate}
  \item A time orientation is needed in order to have a meaningful definition of observers. More precisely, we require that the worldline $x: \mathbb{R} \to M$ of an \underline{observer} have tangent vectors $\dot x_{x(\tau)}h \in C_{x(\tau)}^\#$ for all parameters $\tau \in \mathbb{R}$. This amounts to requiring that observer travel into \underline{future directions} defined by $C^\#$. 
  \item A full definition of observers, which includes not only the above constraint on the tangent vectors of their worldlines, but also a definition of purely spatial directions seen by them, needs to be postponed until we developed more mathematical technology in the next chapter.
  \item Already with the partial definition of observers given above, one may define the energy of a co-vector seen by an observer. More precisely, let $q\in T^*_xM$ and $X\in C^\#$ the tangent vector to an observer's worldline. Then the \underline{(observer-dependent) energy} of the co-vector is $q(X)$. Physically, it will of course only be meaningful to speak of the energy of a co-vector that represents a particle momentum, and we will restrict to those cases later. 
\end{enumerate}

\sep{\bf Definition. } The \underline{cone of positive energy co-vectors} at $x\in M$ is the set
$$(C_x^\#)^+ = \{ p \in T^*_xM  \, |\, p(X)>0 \textrm{ for all } X\in C_x^\#\}\,.$$

\sep{\bf Remarks.}
\begin{enumerate}
  \item Thus at any point $x\in M$ all observers $X\in C^\#_x$ agree on the sign of the energy of any given co-vector $q \in C_x^\#)^+$ at the same point $x\in M$. This is in fact precisely the rationale behind the definition of the cone $C_x^\#$.
  \item If the reduced hyperbolic polynomial $P$ is reducible into factors
$$P(X,k) = P_1(x,k) \cdot \dots \cdot P_f(x,k)\,,$$
then the positive energy cone $(C^\#)^+$ is the sum of the positive energy cones $(C_1^\#)^+, \dots, (C_f^\#)^+$ of the irreducible factors,
$$(C^\#)^+ = (C_1^\#)^+ + \dots +(C_f^\#)^+\,,$$
where the \underline{sum of two subsets} in tangent space is defined as the set of the sum of any two elements of the two sets.    
\end{enumerate}

\sep{\bf Definition.} A bi-hyperbolic polynomial $P_x$ is a called \underline{energy-distinguishing} if the set $N_x = \{k \in T^*_xM \,|\, P(k)=0 \}$ of \underline{massless momenta} is the disjoint union
$$N_x \, = \, N_x^+ \,\, \dot\cup \,\, N_x^-\,,$$
of the set $N^+_x = N \cap (C^\#)^+$ of \underline{positive energy massless momenta} and the set $N^-_x = N \cap -(C^\#)^+$ of \underline{negative energy massless momenta}.

\newpage
\sep{\bf Remarks.}\lift
\begin{enumerate}
\item For matter on geometry $(M,G)$ whose principal polynomial $P_x$ at each point $x\in M$ is an energy-distinguishing and bi-hyperbolic polynomial, all observers agree on the sign of the energy of any massless particle momentum. This allows, for instance, for a unique positive and negative energy split of fundamental solutions if the geometry $G$ is flat.   
\item Energy-distinguishing hyperbolic polynomials are of \underline{even degree}. This is seen as follows. First, one proves that bi-hyperbolicity of $P_x$ implies that 
\begin{equation}\label{posnegcone}
\new{\textrm{closure}((C_x^{\#})^+)\cap -\textrm{closure}((C_x^{\#})^+)=\{0\}\,.}\nonumber
\end{equation} 
Let $k_0$ be such that $k_0\in \textrm{closure}((C_x^{\#})^+)$ and $k_0\in -\textrm{closure}((C_x^{\#})^+)$. It follows from the definition of the dual cone that the following inequalities are true for all $x\in C_x^\#$ : $x.k_0\ge0$ and $x.k_0\le0$. If this were true then the hyperbolicity cone $C_x^\#$ would have to be a plane or a subset of a plane. That would contradict the property of $C_x^\#$ to be open. Second, suppose that the zero set $N_x$ contains a plane. From $\textrm{closure}((C_x^{\#})^+)\cap -\textrm{closure}((C_x^{\#})^+)=\{0\}$ it follows that $(C_x^{\#})^+\setminus \{0\}$ is a proper subset of a half-space. A proper subset of a half-space cannot contain any complete plane through the origin. Hence the \new{existence} of a null plane of $P_x$ would obstruct the energy-distinguishing property. 
Third, this \new{fact} immediately restricts us to cotangent bundle functions $P$ of even degree. For suppose $\deg P$ was odd. Then on the one hand, we would have an odd number of null sheets. On the other hand, the homogeneity of P implies that null sheets in a co-tangent space come in pairs, of which one partner is the point reflection of the other. Together this implies that we would have at least one null hyperplane.
\end{enumerate}

\sep{\bf Definition.} A geometry $(M,G,C^\#)$ is called a \underline{spacetime with respect to matter dynamics} if
\vspace{-.5cm}\begin{enumerate}
\item $P$ is everywhere bi-hyperbolic,
\item $C^\#$ is a time-orientation defined in terms of $P$,
\item $P$ is everywhere energy-distinguishing with respect to the time-orientation,
\end{enumerate}
where $P: T^*xM \to \mathbb{R}$ is the cotangent bundle function defined by the principal polynomial of linear matter dynamics  $D_{MN}(\partial) \Phi^N = 0$ at each point $x\in M$. 

\sep{\bf Remarks.}
\begin{enumerate}
  \item Whether $(M,G)$ presents a viable spacetime structure crucially depends on what linear matter equations one chooses to probe $(M,G)$. To recognize this is not a weakness of the approach presented here, but rather presents a crucial insight.
  \item The restriction to linear matter dynamics is necessary in order to obtain a principal polynomial that depends on the geometry $G$ only, but not on particular solutions of the matter field equations. For while non-linear matter equations may be linearized as $\Phi = \Phi_{exact} + \delta \Phi$ around an exact solution $\Phi_{exact}$, and the principal symbol $P$ of the resulting linearized equations for $\delta \Phi$ still determines the causality of the theory, $P$ will now in general depend on the exact solution $\Phi_{exact}$ around which the theory was linearized. Thus one cannot isolate purely geometric statements if one considers non-linear matter. The lesson is that a spacetime geometry is best probed by linear matter, and thus we focus on such.
  \item Lorentzian geometry $(M,g)$ presents a spacetime with respect to Maxwell theory. For in Lorentz gauge, the Maxwell equations take the form
$$\frac{\delta^n_m}{\sqrt{-\det g(x)}}\partial_a\left(\sqrt{-\det g(x)} g^{ab}(x) \partial_b\, A_n(x)\right) = 0$$
from which one reads off at each $x\in M$ the principal polynomial
$P_x(k) = g^{ab}(x) k_a k_b$, 
which is hyperbolic because the inverse metric $g^{-1}$ has Lorentzian signature. 
The polynomial $P^\#(X) := g_{ab} X^a X^b$ is then indeed dual to $P$, since
$$P^\#(DP(k)) = g_{ab}(2 g^{am} k_m) (2 g^{bn} k_n) = g^{mn} k_n k_n = 0$$
for all $k$ with $P(k)=0$, and also hyperbolic since $g$ has Lorentzian signature.
Thus $P$ is b-hyperbolic. It is also energy-distinguishing, since choosing one of the two hyperbolicity cones of $P^\#$ as the time-orientation $C^\#$, one finds that
$$(C^\#)^+ = \{ q \in T_x^*M \, | \, q(X)>0 \textrm{ for all } X\in C^\# \} = \textrm{closure}(C)$$
where $C$ is the hyperbolicity cones of $P$ that lies entirely within $(C^\#)^+$. But then $N = N^+ \, \dot\cup\, N^-$ and $P$ is also energy-distinguishing. 
\end{enumerate}

\newpageja
\section*{Lecture VI: \, MASSIVE DISPERSION}
Not only for completeness, but because of its crucial role in mapping massive co-vectors to associated velocity vectors, we study massive point particles on spacetimes in this lecture. An important corollary is a meaningful classification of all physically viable modified dispersion relations. 
   
\sep{\bf Definition.} Let $C\subset (C^\#)^+$ be a hyperbolicity cone of positive energy. Then any $q\in C$ is called a \underline{massive positive energy momentum}, and its \underline{mass} $m>0$ is given by $P(q)=m^{\deg P}$.

\sep{\bf Remarks.}
\begin{enumerate}
  \item There is always one hyperbolicity cone $C$ of $P$ that lies entirely within $(C^\#)^+$. For we know that the boundary $\partial C$ of any hyperbolicity cone is a connected set of $P$-null co-vectors, and the disjoint union $(C^\#)^+ \dot\cup - (C^\#)^+$ covers the entire set of null co-vectors. Hence either $C$ or $-C$ lies entirely within the positive energy cone $(C^\#)^+$.  
\item The positivity of $P$ on the cone $C$ of massive positive energy momenta is guaranteed by our choice of the overall sign in the definition of $P$, in lecture III. 
\end{enumerate} 

\sep{\bf Theorem.} For a bi-hyperbolic and energy distinguishing $P$, the so-called \underline{barrier function}
$$f_x: C_x \to \mathbb{R}, \qquad f_x(q) = -\frac{1}{\deg P} \ln P(x,q)$$
is strictly convex and essentially smooth, which guarantees that the 
\underline{Legendre map}
$$L_x: C_x \to L_x(C_x) \subset T_xM, \qquad L_x(q) := - D(\ln P)(x,q)$$
is invertible, with 
$$L_x^{-1}: L_x(C_x) \to C_x, \qquad L_x^{-1}(q) = - Df_x^L$$
given in terms of the \underline{Legendre transform}
$$f_x^L: L_x(C_x) \to \mathbb{R}, \qquad f_x^L(X) = - L_x^{-1}(X)X - f_x(L_x^{-1}(X))$$
of the barrier function $f_x$, at each point $x$ of the manifold $M$.

\newpage
\sep{\bf Remarks.}
\begin{enumerate}
  \item Essential smoothness refers to a particular behaviour of the barrier function close to the boundary of the convex set on which it is defined; see any text on convex analysis. The functions to be Legendre transformed in classical mechanics or thermodynamics are usually defined on an entire vector space, which is of course a convex set but has no boundary, so that the criterion of essential smoothness is trivially satisfied there, and thus less known.
  \item It is only the interplay of bi-hyperbolicity and the energy-distinguishing property that ensures the existence and invertibility of the above Legendre map. Thus all criteria for a spacetime structure, as laid out at the end of the previous chapter, are required to have the present Legendre theory at our disposal, which in turn will now be used to derive the dynamics of massive particles. 
\end{enumerate} 

\sep{\bf Theorem.} The \underline{dynamics of a free positive energy particle of mass} $m>0$ is encoded in the action
$$S[x] = m \int d\tau \, P_{x(\tau)}^*(\dot x(\tau))^{1/{\deg P}}\,,$$
where at each point $x\in M$, the function $P_x^*(X) = P_x(L_x^{-1}(X))^{-1}$ is defined on the entire cone $L_x(C_x)$ in tangent space. 

\sep{\it Proof.} It is obvious that the Helmholtz action 
$$S[x,q,\lambda] = \int d\tau\, \left[q_a \dot x^a - \lambda m \ln P(x,\frac{q}{m})\right]$$
leads, upon variation with respect to $\lambda$, to the dispersion relation for a positive energy particle momentum $q$ of mass $m$, and together with the geometry-free term $q_a \dot x^a$ presents the action for such a particle, very similar to the massless case discussed in lecture IV. Variation with respect to $q$ then yields $\dot x^a = (\lambda \deg P) L_x(q/m)$, 
which we know how to invert due to the Legendre theorem given above, so that
$$q = m L_x^{-1}(\frac{\dot x}{\lambda \deg P})\,.$$ 
Using the thus given relation between massive momenta and the tangent vector to the particle worldline, as well as the definitions of the barrier function and the Legendre dual, one eliminates $q$ from the action and obtains 
\begin{equation}\label{intermediate}
  S[x,\lambda] \new{= - m \deg P \int d\tau \, \lambda f^L(\dot x/(\lambda \deg P))} = -m \deg P\int d\tau \left[\lambda f_x^L(\dot x) + \lambda \ln(\lambda\deg P)\right]\,,\nonumber
\end{equation}
where for the second equality we used the easily verified scaling property $f^L(\alpha \dot x) = f^L(\dot x) - \ln \alpha$. From variation of this equivalent action with respect to $\lambda$, one then learns that 
\begin{equation}
  f^L(\dot x) + \ln(\lambda \deg P) + 1 = 0\,.\nonumber
\end{equation}
Using this twice, one has $\lambda  f_x^L(\dot x) +\lambda \ln (\lambda \deg P))= -\lambda = - \exp(-f_x^L(\dot x)-1)/\deg P$. Noting that because of $\dot x\in L_x(C_x)$ one also has $L^{-1}(x,\new{\dot x})(\dot x) = 1$ and thus $f_x^L(\dot x)=-1-f_x(L^{-1}(\dot x))$, one eliminates $\lambda$ to finally arrive at the equivalent action
\begin{equation}
\label{finsleraction}
  S[x] =m \int d\tau P^*(x,\dot x)^{1/\deg P}\nonumber
\end{equation}
for a free point particle of positive mass $m$. This concludes the proof. 
 
\sep{\bf Remarks.}
\begin{enumerate}
\item While the tangent bundle function $P^*$ is generically non-polynomial, it is elementary to see that it is homogeneous of degree $\deg P$.
\item 
The action we arrived at above is \underline{reparametrization invariant}, as it should be. However, parametrizations for which $P(x,L^{-1}(x,\dot x)) = 1$ along the curve are distinguished since they yield the simple relation 
\begin{equation}\label{simplemomentum}
  \dot x = L_x(q/m)\nonumber
\end{equation}
between the free massive particle velocity $\dot x$ and the particle momentum $q$ everywhere along the trajectory $x$. As usual, we choose such clocks and call the time they show proper time.
  \item
 Thus the \underline{physical meaning} of the Legendre map is established (namely mapping massive positive energy particle momenta to the respective tangent vectors of their worldlines), and one may thus justifiably call the open convex cone $L_x(C_x)$ the cone of massive particle velocities, and the function $P^*$ the massive dual of $P$, which indeed encodes the tangent bundle geometry seen by massive particles.
\item 
Reassuringly, one can now prove that the observer cone lies in the massive dual,  $C_x^\# \subseteq L_x(C_x)$. Thus one may think of observers as massive, as usual. The converse, however, does not hold, since the inclusion is generically proper.
\end{enumerate}

\sep{\bf Definition.} An \underline{observer} is a curve $e: \mathbb{R} \to LM$ in the frame bundle $LM$ over $M$ such that 
\begin{enumerate}
  \item the first frame vector $e_0$ at each point coincides with the tangent vector to the canonically projected curve $\pi \circ e: \mathbb{R} \to M$,
  \item $e_0$ at each point of the curve lies within the observer cone $C^\#$ and
  \item $L^{-1}(e_0)(e_\alpha) = 0$ for the remaining frame vectors $e_\alpha$ with $\alpha=1, \dots, \dim M - 1$.
\end{enumerate} 

\sep{\bf Remarks.}
\begin{enumerate}
 \item The vector subspace $V_x = \{X \in T_xM \, | \, L^{-1}(e_0)(X)=0 \}$ contains the \underline{purely spatial} directions seen by the observer. 
 \item An observer frame in each tangent space along the curve $\pi \circ e$ induces a unique dual frame
$$\epsilon^0 = L^{-1}(e_0), \epsilon^{1}, \dots, \epsilon^{\dim M - 1}\,,$$
and the zero component $q_0$ of a co-vector $q = q_0 \epsilon^0 + q_\alpha \epsilon^\alpha$, with a sum over $\alpha=1, \dots, \dim M - 1$ understood, coincides precisely with the energy of that co-vector as seen by the observer.     
\end{enumerate}

\sep{\bf Theorem.} The \underline{dispersion relation} of causally propagating matter of mass $m> 0$, whose sign of energy is agreed upon by all observers, must take the form 
$$P(E \epsilon^0 + p_\alpha \epsilon^\alpha) = m^2$$
for a bi-hyperbolic and energy-distinguishing $P$, where $\epsilon^0, \dots \epsilon^{\dim M - 1}$ is an observer co-frame. Solving the above relation for $E$ then yields the relation between the energy $E$ and the purely spatial momentum $p_\alpha \epsilon^\alpha$ seen by the particular observer chosen in the decomposition of the massive positive energy momentum $p \in C$. 

\newpage
\sep{\bf Remarks.}
\begin{enumerate}
  \item Likewise one obtains the physically meaningful massless dispersion relations, by letting $p \in N^+$ and $m=0$.
  \item Note that the polynomial $P$ enters into the above decomposition twice: firstly in determining what constitutes, for a particular observer, a split into the purely temporal and purely spatial directions; and secondly in determining the dispersion relation as such. In order to play either of these roles, $P$ must be bi-hyperbolic and energy-distinguishing, as we saw in these lectures. 
  \item It is obvious that once the above relation has been solved for $E$ in terms of the purely spatial components of the momentum, to take the form
$$E = \sum_{i=0}^\infty c^{\alpha_1 \dots \alpha_i} p_{\alpha_1} \dots p_{\alpha_i}, \qquad \textrm{ with } c = m\,,$$ 
it is prohibitively difficult to recognize from the generically infinitely many coefficients $c, c^\alpha, c^{\alpha\beta}, \dots$, whether the underlying polynomial $P$ was bi-hyperbolic and energy-distinguishing (and thus whether the dispersion relation is one of causally propagating matter of definite sign of energy at all). But even if that were so, it appears rather questionable to attempt to bound the coefficients from comparison with experiment without understanding the structure of observer frames, for then the quantities $E$ and $p_\alpha \epsilon^\alpha$ are void of any meaning. Corresponding attempts in the phenomenology literature are thus to be severely doubted. 
\end{enumerate}

\newpageja
\section*{Lecture VII. SUPERLUMINALITY}
Superluminal motion of massive particles can occur in all spacetimes with $\deg P > 2$. This is compatible with causality by construction in our spacetimes, but raises the question of what mechanism prevents ordinary observation of superluminal massive particles. In this lecture we find the answer to this question in the fact that superluminal massive particles are kinematically allowed to radiate off massless particles until they are infraluminal. 

\sep{\bf Theorem.} The process where a positive energy massive particle of momentum $p$ radiates off a positive energy massless particle at a point $x$ in spacetime 
is \underline{kinematically forbidden} if and only if $p$ lies in the cone 
\begin{equation}\label{stabco}
  L_x^{-1}(C_x^\#) \subseteq C\,.
\end{equation}
in cotangent space. 

\sep{\bf Remarks.}\lift
\begin{enumerate}
  \item The cone $L_x^{-1}(C_x^\#)$ contains precisely those positive energy massive momenta that correspond to \underline{infraluminal} massive particles, since the slowest light is the one on the boundary of $C^\#$. All other positive energy massive particles travel at a velocity higher than the slowest light. Depending on the geometry, some may travel even faster than the fastest light. 
  \item For the familiar Lorentzian metric case ($\deg P = 2$), one of course obtains that $L_x^{-1}(C_x^\#) = C_x$; in other words, positive energy massive particles travel at a velocity lower than the speed of light and cannot radiate off a massless particle in vacuo. This is often stated as that there is no vacuum Cerenkov radiation, which we now see is only true for Lorentzian spacetimes. 
  \item The proof of the theorem is given in [A].
\end{enumerate}

\enlargethispage{1cm}
\sep{\bf Reduction to 1+1 dimensions.} In order to understand what precisely lies in store with the superluminal particles, without obscuring the essentially straightforward argument by cumbersome algebra, we consider the situation in $1+1$ spacetime dimensions. Extension to the physically relevant $3+1$ dimensions presents no conceptual challenges. 

\begin{figure}[h]
\includegraphics[width=10cm,angle=0]{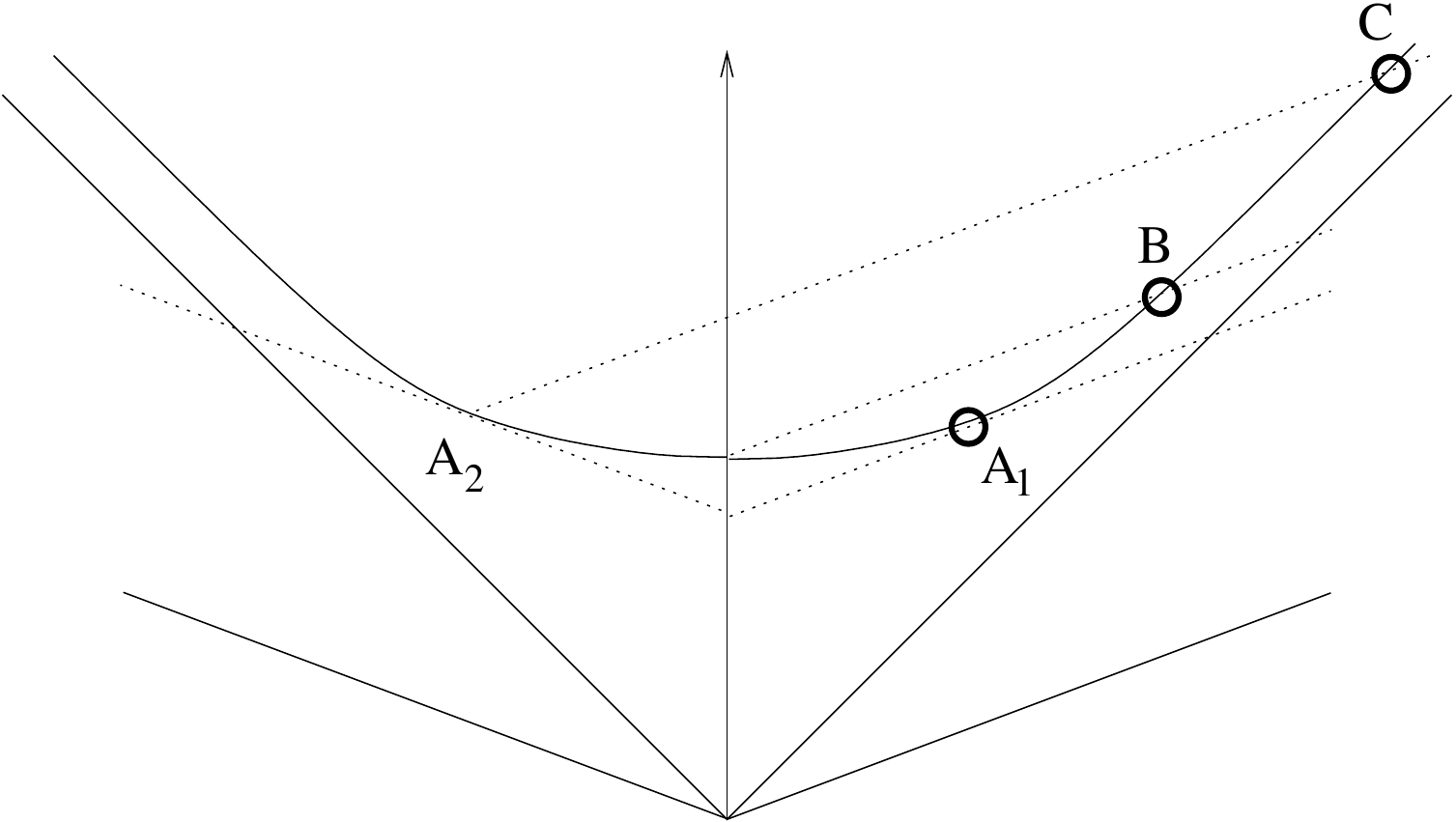}
\caption{\label{fig_points} Massless momenta, mass shell and construction of distinguished points thereon}
\end{figure}

Figure \ref{fig_points} shows the massless momenta in some cotangent space, together with the mass shell for mass $m$. Note that (because the Gauss map sends `inner' massless cones in cotangent space to outer cones in tangent space, and vice versa) the inner cone here corresponds to fast massless particles, while the outer one corresponds to slow massless particles. 

\begin{figure}[h]
\includegraphics[width=10cm,angle=0]{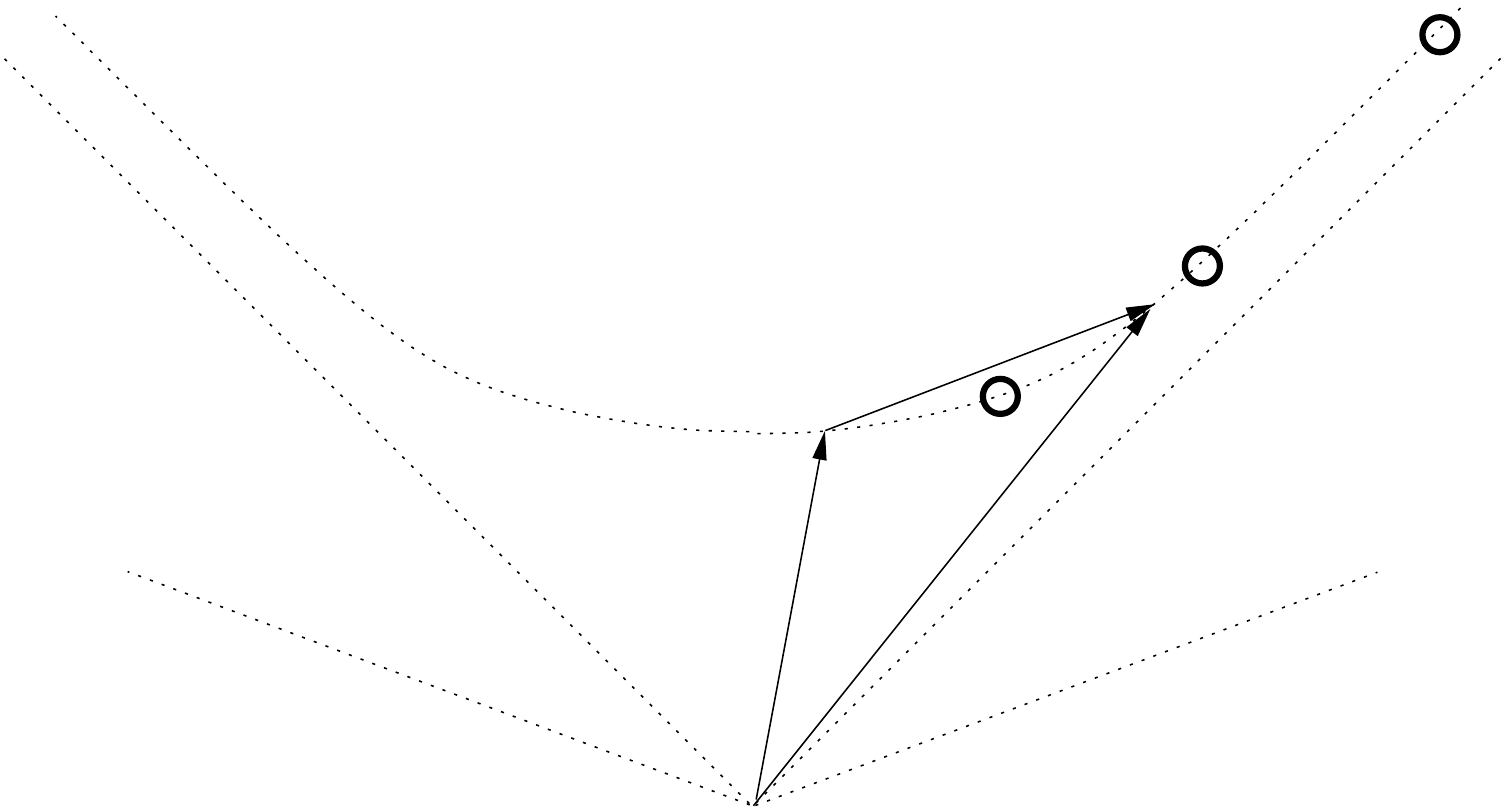}
\caption{\label{fig_decay} Momentum diagram for a massive particle radiating off one slow massless particle.}
\end{figure}

Now we construct the points $A_1$ and $A_2$ as those points on the mass shell, where the slow massless momentum cone touches the mass shell. The relevance of these points is that any massive momentum of higher energy can radiate off a massless particle traveling at the speed of slow massless particles, see figure \ref{fig_decay}. Thus massive particles whose momenta on the mass shell lie between the points $A_1$ and $A_2$ are precisely those that cannot decay. Thus the straight lines connecting the origin with $A_1$ and $A_2$, respectively, must constitute the boundary of the cone $L^{-1}(C^\#)$ identified above. 

Now construct the point $C$ as the intersection of the slow massless cone centered at $A_2$ with the mass shell. Obviously any momentum on the mass shell that lies between the points $A_1$ and $C$ can radiate off at most one massless particle, since the outgoing massive particle momentum will then lie between $A_2$ and $A_1$ and is thus kinematically protected form further decay. Of course a similar point (not shown) could be constructed on the left half of the mass shell. 

Finally we obtain the point $B$ (and a similar point on the left, not shown) on the mass shell as the intersection of the slow massless momentum cone centered at the point where the massive particle energy is the lowest with the mass shell. The significance of this point is that it marks the point where the outgoing massive particle ceases to still run into the same direction as the ingoing massive particle, since for any ingoing massive momentum between the points $B$ and $C$ the outgoing particle momentum is going into the opposite direction.      

\begin{figure}[h]
\includegraphics[width=14cm,angle=0]{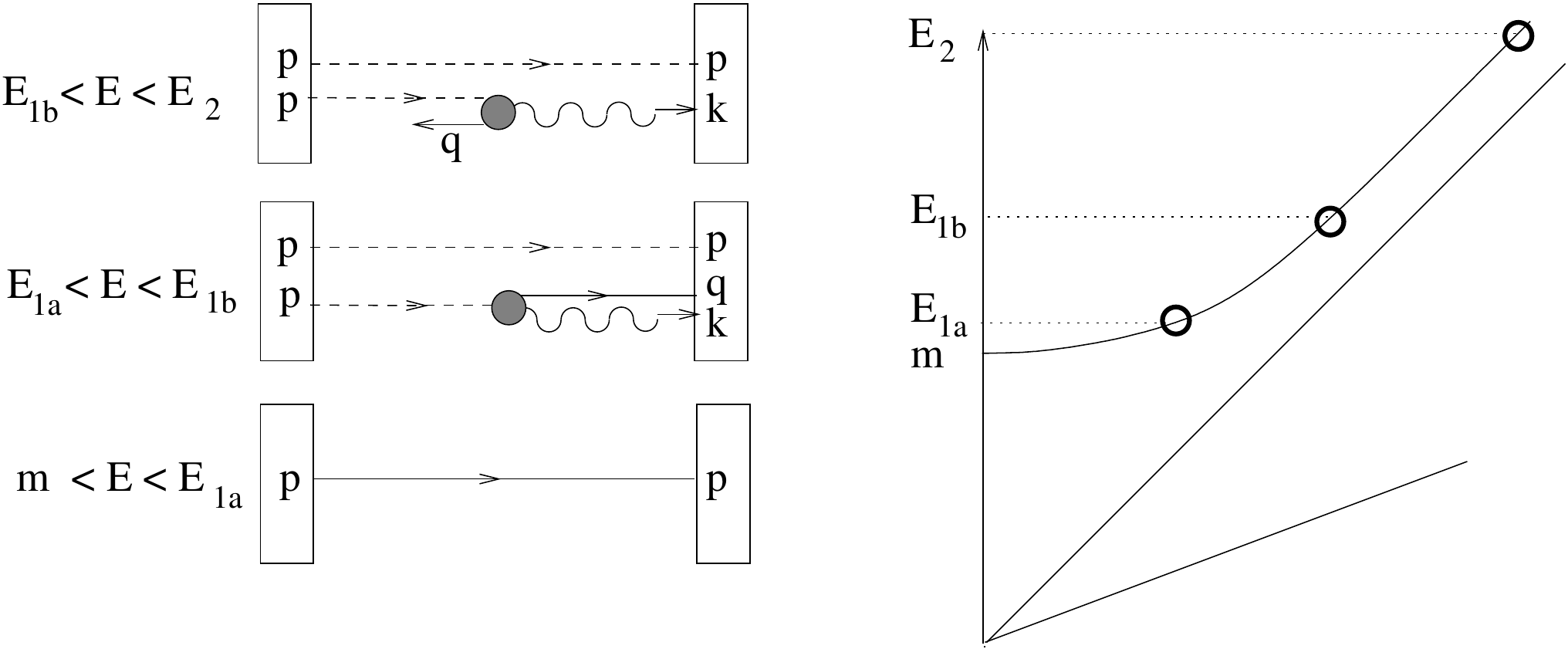}
\caption{\label{fig_arriving} Particles emitted from a source (left) at certain energy arriving at a detector (right); dashed lines are superluminal massive, solid lines infraluminal massive, wiggly lines slow massless}
\end{figure}

This leaves us with the following expected \underline{pattern of arrival} of superluminal and infraluminal particles at some detector if the emitted massive particles at some source had a momentum $p$, see figure \ref{fig_arriving}. For the energy range $m < E < E_{1a}$, between the rest mass $m$ of the emitted particle and the energy $E_1$ above which radiating off a massless particle becomes possible, the particle is stable and thus arrives at the detector with exactly the momentum it was emitted with. For the energy range $E_{1a} < E < E_{1b}$, where particles are necessarily superluminal and $E_{1b}$ was constructed to be the energy above which the massive particle no longer travels in the same direction after it emitted the massless particle, one either detects the superluminal particle with unchanged momentum (because no massless particle was actually radiated off) or an infraluminal massive particle whose momentum is obtained from the construction in figure \ref{fig_decay} as a function of the emitted particle momentum. In the energy region $E_{1b} < E < E_2$, where $E_2$ was constructed as the highest energy a superluminal particle can have in order to radiate off at most one massless particle, one however expects at most superluminal particles to arrive at the detector at precisely the momentum they were emitted with. For if a superluminal particle in this energy range were to radiate, its momentum would be reversed but then lie in the stable energy region $m < E < E_{1a}$. Thus the prediction from this simple model in $1+1$ dimensions is that there is an energy range within which no  superluminal particles are detected.    

\enlargethispage{1cm} Calculating the \underline{quantum mechanical decay rate} on tensorial spacetimes is possible, but beyond the scope of these lectures. 

\newpageja
\section*{Lecture VIII: \, SPACETIME DYNAMICS}
The aim of this final lecture is to find dynamics that develop initial geometric data from one $P$-spacelike hypersurfaces to another, such that sweeping out the spacetime manifold this way, one reconstructs a bi-hyperbolic and energy-distinguishing dispersion relation everywhere. To this end, one studies hypersurfaces with hyperbolic co-normals by their embedding maps and thus calculates how functionals of this embedding map change under normal and tangential deformations of the hypersurface. This change can be expressed by a linear action of deformation operators on such functionals, and it is the commutation algebra of these deformation operators that needs to be represented on the phase space of spatial geometries in order to obtain canonical dynamics for the $(\deg P, 0)$-tensor field $P$. 

\sep{\bf Definition.} Let $X: \Sigma \hookrightarrow M$ be a smooth embedding map of a smooth manifold $\Sigma$ of dimension $\dim M - 1$ with local coordinates $\{y^\alpha\}$ into the smooth manifold $M$ with coordinates $\{x^a\}$. Then the $\dim M$ vectors
$$T(y) := L_{X(y)}(n(y)), \quad e_1(y) := \frac{\partial X^a(y)}{\partial y^1} \frac{\partial}{\partial x^a}, \quad\dots,\quad e_{\dim M-1}(y) := \frac{\partial X^a(y)}{\partial y^{\dim M - 1}} \frac{\partial}{\partial x^a} $$
constitute a basis of each tangent space $T_{X(y)}M$ of $M$ along the hypersurface $\Sigma$, where the co-normals $n(y)$ are uniquely determined by the conditions
$$n \in C, \qquad \qquad P(n) = 1, \qquad n(e_\alpha)=0 \textrm{ for all } \alpha=1,\dots,\dim M - 1\,.$$
The first basis vector $L_{X(y)}(n(y))$ is called the \underline{(spacetime) normal vector} to the hypersurface $X(\Sigma)$ at each point $y\in \Sigma$, while the remaining basis vectors $e_1(y), \dotsm e_{\dim M - 1}(y)$ are \underline{(spacetime) tangent vectors} to $X(\Sigma)$. The uniquely determined \underline{co-basis} be denoted 
$$n(y), \quad \epsilon^1(y),\quad  \dots,\quad \epsilon^{\dim M-1}(y)\,.$$

\sep{\bf Remarks.}
\begin{enumerate}
  \item The spacetime geometry explicitly enters into the above definitions only in 
determining the hypersurface normal vector $L(n)$, but there three-fold: first in determining the hyperbolicity cone $C$ to ensure that the hypersurface in question can be an initial data surface, second in normalizing the canonically defined co-normal which is otherwise only determined up to positive rescalings, and third in mapping the co-normal to a normal vector by virtue of the Legendre map $L$.
  \item The embedding map determines the \underline{purely spatial geometry} 
$$P^{\alpha_1 \dots \alpha_I}(y)[X] := P(\underbrace{\epsilon^{\alpha_1}(y), \dots, \epsilon^{\alpha_I}(y)}_{I}, \underbrace{n(y), \dots, n(y)}_{\deg P - I})\,,\qquad I=2,\dots,\deg P\,$$
on the hypersurface. At least this is the purely spatial geometry \underline{detected by point particles}. 
\item One may instead also construct the \underline{purely spatial geometry seen by fields}, by considering all non-vanishing hypersurface tensor fields obtained by inserting the  $\epsilon^{\alpha}$ and $n$ into the co-vector slots and $e_\alpha$ and $T$ into the vector slots of the geometric tensor $G$. The precise number and nature of the fields then depends on the precise structure (valence and algebraic symmetries) of $G$. We will thus focus in the following on the geometry seen by point particles, but the principle is the same for the geometry seen by fields. 
For an area metric geometry ($G^{abcd} = G^{cdab} = -G^{bacd}$), for instance, one obtains in $d$ spacetime dimensions the three purely spatial tensor fields 
\begin{eqnarray}
G^{\alpha_1\alpha_2}(y)[X] &=& G(n(y),\epsilon^{\alpha_1}(y),n(y),\epsilon^{\alpha_2}(y))\,,\nonumber\\
G^{\alpha_1\alpha_2\alpha_3}(y)[X] &=& G(n(y),\epsilon^{\alpha_1}(y),\epsilon^{\alpha_2}(y),\epsilon^{\alpha_3}(y))\,,\nonumber\\ G^{\alpha_1\alpha_2\alpha_3\alpha_4}(y)[X] &=& G(\epsilon^{\alpha_1}(y),\epsilon^{\alpha_2}(y),\epsilon^{\alpha_3}(y),\epsilon^{\alpha_4}(y))\,.\nonumber
\end{eqnarray}
\item In any case, note that the quantities defined under 2. and 3.,  describing the purely spatial geometry, are (i) tensors on $\Sigma$ and (ii) \underline{functionals} of the embedding map. While similar quantities are defined for $I=0$ and $I=1$, they are constantly $1$ and $0$, respectively, due to our definition of $n$.  
\item Assume we are already given a spacetime $(M,G,C^\#)$ with an associated cotangent bundle $P$, and consider a hypersurface embedding $X: \Sigma \hookrightarrow M$. Then one can study how functionals of $X$, such as the geometric data in the previous remark, change under deformations of the hypersurface. Such deformations in normal and tangential directions of the hypersurface are encoded by specifying a hypersurface-scalar field $N$ and hypersurface-vector field $N^\alpha e_\alpha$, respectively. The change of a functional $F$ of $X$ is then given, to linear order, by $\mathcal{H}(N) F$ and $\mathcal{D}(N^\alpha e_\alpha)F$, where
$$\mathcal{H}(N) = \int_\Sigma d^{d-1}y \, N(y) T^a(y) \frac{\delta}{\delta X^a(y)} \quad \textrm{ and } \quad  \mathcal{D}(N^\alpha e_\alpha) = \int_\Sigma d^{d-1}y \, N^\alpha(y) e^a_\alpha(y) \frac{\delta}{\delta X^a(y)}$$
are the \underline{normal} and \underline{tangential deformation operator}, respectively.
\end{enumerate}

\newpage
\sep{\bf Theorem.} The normal and tangential deformation operators, acting on functionals on an initial data hypersurface, satisfy the \underline{hypersurface deformation algebra}
\begin{eqnarray}
  {[\mathcal{H}(N),\mathcal{H}(M)]} &=& -\mathcal{D}((\deg P-1) P^{\alpha\beta}(M \partial_\beta N - N \partial_\beta M)\partial_\alpha)\,,\nonumber\\
  {[\mathcal{D}(N^\alpha \partial_\alpha),\mathcal{H}(M)]} &=& - \mathcal{H}(N^\alpha \partial_\alpha M)\label{DH}\,,\nonumber\\
  {[\mathcal{D}(N^\alpha\partial_\alpha),\mathcal{D}(M^\beta \partial_\beta)]} &=& - \mathcal{D}((N^\beta \partial_\beta M^\alpha - M^\beta \partial_\beta N^\alpha)\partial_\alpha)\,.\nonumber
\end{eqnarray}

\sep{\bf Remarks.}
\begin{enumerate}
  \item Note that the background geometry enters the hypersurface deformation algebra exclusively through the definition of the normal vectors $T(y)$ to each point $X(y)$ of the hypersurface $X(\Sigma)$, since these are obtained from the canonically defined normal co-vectors $n(y)$ by virtue of the Legendre map $L$, which indeed is determined through $P$. Also observe that only the first commutation relation depends on the spacetime geometry, while the other two are fully independent of it.
  \item The hypersurface deformation operators are constructed such as to describe the change of functionals if one moves from one initial value hypersurface to another one near-by, if indeed the entire spacetime geometry is already known. Now the problem of finding \underline{dynamics for the spatial geometry} is to find equations that provide the spatial geometry on a near-by hypersurface solely from the spatial geometry on the original hypersurface, without the entire spacetime geometry already being known. This is of course at most possible if one compensates for the lack of information about the geometry around the hypersurface by prescribing additional initial data on the hypersurface, in form of \underline{canonical momenta} 
$$\hat\pi_{\alpha_1\alpha_2}, \quad \hat\pi_{\alpha_1\alpha_2\alpha_3}, \quad \dots, \quad \hat\pi_{\alpha_1 \alpha_2 \dots \alpha_{\deg P}}$$ 
associated with the \underline{purely spatial geometric data} given by
$$\hat P^{\alpha_1\alpha_2}, \quad \hat P^{\alpha_1\alpha_2\alpha_3}, \quad \dots, \quad \hat P^{\alpha_1 \alpha_2 \dots \alpha_{\deg P}}\,,$$
where these are now no longer understood as induced from a known spacetime geometry, but as independent tensors on $\Sigma$; to avoid any conceptual confusion, we marked these objects with a hat. The space of the tensor fields $(P^A, \pi_A)$ is called the \underline{geometric phase space}, and the requirement that the $\pi_A$ be canonically conjugate to the $P^A$ fixes the \underline{Poisson bracket} on geometric phase space to
$$\{\hat F, \hat G\} = \int_\Sigma dy \, \left[\frac{\delta\hat F}{\delta \hat{P}^A(y)}\frac{\delta \hat G}{\delta \hat{\pi}_A(y)} - \frac{\delta\hat G}{\delta \hat{P}^A(y)}\frac{\delta \hat F}{\delta \hat{\pi}_A(y)}\right]\,,$$  
where the observables $\hat F$ and $\hat G$ are functionals of the phase space variables, and a capital Latin index $A$ denotes the entire collection of indices $(\alpha_1\alpha_2, \alpha_1\alpha_2\alpha_3, \dots, \alpha_1\alpha_2\dots\alpha_{\deg P})$ and summation over a capital latin index is summation over all those spacetime indices.
\item As discussed before, one may alternatively wish to construct the phase space of the purely spatial geometry seen by fields. The number of fields replacing the $P^{\alpha_1\alpha_2}, \dots P^{\alpha_1 \alpha_2 \dots \alpha_I}$ in the previous construction then depends on the precise structure of the geometric tensor $G$. For an area metric geometry ($G^{abcd} = G^{cdab} = -G^{bacd}$) in $d$ spacetime dimensions, the initial data on a hypersurface are described by the three tensor fields 
$$\hat G^{\alpha_1\alpha_2}, \quad\hat G^{\alpha_1\alpha_2\alpha_3}, \quad \hat G^{\alpha_1\alpha_2\alpha_3\alpha_4}$$
together with canonically conjugate momenta
$$\hat\gamma_{\alpha_1\alpha_2}, \quad \hat\gamma_{\alpha_2\alpha_2\alpha_3}, \quad \hat\gamma_{\alpha_1\alpha_2\alpha_3\alpha_4}\,.$$
The Poisson brackets are constructed accordingly.
\end{enumerate}

\sep{\bf Definition.} Dynamics for observables on the geometric phase space is given by the \underline{Hamiltonian}
$$H = \int_\Sigma d^{d-1}y \,\, \left[\hat{\mathcal{H}}(N) + \hat{\mathcal{D}}(N^\alpha e_\alpha)\right]\,,$$
where the \underline{superhamiltonian} $\hat{\mathcal{H}}$ and \underline{supermomentum} $\hat{\mathcal{D}}$ are \underline{functionals of the geometric} \underline{phase space variables} $(P^A, \pi_A)$ that represent the hypersurface deformation algebra by virtue of
\begin{eqnarray}
  \{\hat{\mathcal{H}}(A),\hat{\mathcal{H}}(B)\} &=& \hat{\mathcal{D}}((\deg P-1) \hat{P}^{\alpha\beta}(A \partial_\beta B - B \partial_\beta A)\partial_\alpha)\,,\nonumber\\
  \{\hat{\mathcal{D}}(A^\alpha \partial_\alpha),\hat{\mathcal{H}}(B)\} &=&  \hat{\mathcal{H}}(A^\alpha \partial_\alpha B)\,,\nonumber\\
  \{\hat{\mathcal{D}}(A^\alpha\partial_\alpha),\hat{\mathcal{D}}(B^\beta \partial_\beta)\} &=&  \hat{\mathcal{D}}((A^\beta \partial_\beta B^\alpha - B^\beta \partial_\beta A^\alpha)\partial_\alpha)\,.\nonumber
 \end{eqnarray}

\newpage
\sep{\bf Remarks.}
\begin{enumerate}
  \item This definition describes precisely what dynamics is all about: geometric data on an initial data hypersurface are evolved such that, if put on a neighbouring hypersurface (determined from the original hypersurface by the deformation fields $N$ and $N^\alpha e_\alpha$), and this data then again put on a neighbouring hypersurface, and so forth, the entirety of spatial geometries on all these hypersurfaces constitutes an entire spacetime geometry. The representation requirement then ensures that this is consistent with the idea that the the data on the individual hypersurfaces are really nothing but the purely spatial geometries induced from that overall spacetime geometry, with all its required properties.  
  \item With the mathematical technology for (generalized) spacetime geometries developed in these lectures, it can be shown from the third Poisson bracket relation that the \underline{supermomentum} for the geometry seen by point particles is uniquely given by 
$$ \hat{\mathcal{D}}(\vec{N}) = \sum_{I=2}^{\deg P} \int_\Sigma dy\, N^\beta(y) \left[\partial_\beta \hat P^{\alpha_1 \dots \alpha_I} \hat \pi_{\alpha_1 \dots \alpha_I} + I\, \partial_{\alpha_1}(\hat P^{\alpha_1 \dots \alpha_I}\hat  \pi_{\alpha_2 \dots \alpha_I \beta})\right]\,.$$
  \item From the second Poisson bracket relation it is possible to show that the localized superhamiltonian $\hat{\mathcal{H}}(y) := \hat{\mathcal{H}}(\delta_y)$ for the geometry seen by point particles is, first, a scalar density of weight one, and, second, decomposes into \underline{non-local} and \underline{local} parts according to
$$\hat{\mathcal{H}}(y)[\hat P,\hat \pi] = \hat{\mathcal{H}}_\textrm{local}(y)[\hat P](\hat \pi) + \hat{\mathcal{H}}_\textrm{non-local}(y)[\hat P,\hat \pi]\,,$$
of which the non-local part however is directly determined to be
$$\hat{\mathcal{H}}_{\textrm{non-local}}(y) = \sum_{I=2}^{\deg P}\big[(\deg P-I)\partial_{\beta}(\hat P^{\beta\alpha_1\dots\alpha_I}\hat \pi_{\alpha_1\dots\alpha_I}) - (\deg P-1)\,I\,\partial_{\beta}(\hat P^{\alpha_2\dots\alpha_I}\hat P^{\alpha_1\beta} \hat \pi_{\alpha_1\dots\alpha_I})\big](y)\,,$$
and only the local part $\hat{\mathcal{H}}_\textrm{local}$ remains to be determined from the first Poisson bracket.
\item With the supermomentum and non-local part of the superhamiltonian already known, the first Poisson bracket relation finally presents a condition quadratic in the local part of the superhamiltonian. Remarkably, this is equivalent to a linear condition in the Legendre transform of $\hat{\mathcal{H}}_\textrm{local}$ in the $\hat \pi^A$ variables,
$$L(y)[\hat P](K) := \hat\pi_A(y)[\hat P](K) K^A(y) - \hat{\mathcal{H}}(y)_\textrm{local}[\hat P](\hat\pi[\hat P](K))\,,$$
where the Legendre dual variables $K^A$ are given by
$$K^A(y) := \frac{\partial \hat{\mathcal{H}}(y)_\textrm{local}}{\partial \pi_A(y)}\,.$$
For with these definitions, the remaining first Poisson bracket relation takes the form of the linear functional differential equation
\begin{eqnarray}\nonumber
   0 &=& - \frac{\delta L(x)}{\delta \hat{P}^A(y)} K^A(y) + \partial_{y^\zeta}\left[\frac{\delta L(x)}{\delta \hat{P}^A(y)} M^{A\zeta}(y)\right] - \frac{\partial L(x)}{\partial K^A(x)}K^B(x)Q_{B}{}^{A\beta}(x)\partial_\beta\delta_x(y)\nonumber\\
&&+\frac{\partial L(x)}{\partial K^A(x)}\left[ R^{A\mu\nu}(x)\partial^2_{\mu\nu} \delta_x(y) - S^{A\mu}(x) \partial_\mu \delta_x(y) \right] - (x \leftrightarrow y)\nonumber 
\end{eqnarray}
where the coefficients $R^{A\mu\nu}$, $Q_A{}^{B\mu}$ and $M^{A\beta}$ contain only the configuration variables, but not their derivatives,
\begin{eqnarray}
 R^{\alpha_1\dots\alpha_I\mu\nu} &=& I(\deg P\min1)P^{(\beta|(\alpha_1}P^{\alpha_2\dots\alpha_I)|\mu)}\,,\nonumber\\
 M^{\alpha_1\dots\alpha_I\, \beta} &=&-(\deg P \min I) \hat{P}^{\beta\alpha_1\dots\alpha_I} + I (\deg P\min 1) \hat{P}^{\beta(\alpha_1} \hat{P}^{\alpha_2\dots\alpha_I)}\,,\nonumber\\
Q_{\alpha_1\dots\alpha_K}{}^{\beta_1\dots\beta_I\, \mu} &=& \delta^K_{I+1} (\deg P \min I) \delta^{\mu \beta_1\dots\beta_I}_{(\alpha_1\dots\alpha_{I+1})}
 - \delta^K_2 I (\deg P\min 1) \hat{P}^{(\beta_2\dots\beta_I} \delta^{\beta_1)\mu}_{(\alpha_1\alpha_2)}\nonumber\\
 &&- \delta^K_{I-1} I (\deg P \min 1) \hat{P}^{\mu(\beta_1}\delta^{\beta_2\dots\beta_I)}_{\alpha_1\dots\alpha_{I-1}}\,,\nonumber
\end{eqnarray}
but the coefficients $S^{A\mu}$ also containing their first partial derivatives,
\begin{eqnarray}\nonumber
S^{\alpha_1\dots\alpha_I\mu}&=&I(\deg P\min1)(\deg P\min2)P^{\mu\gamma(\alpha_1}\partial_\gamma P^{\alpha_2\dots\alpha_I)}+I(\deg P \min1)(\deg P\min I)\partial_\gamma P^{\gamma(\alpha_1}P^{\alpha_2\dots\alpha_I)\mu}\nonumber\\
&&+(\deg P\min 1)P^{\mu\gamma}\partial_\gamma P^{\alpha_1\dots\alpha_I}-I(I\min1)(\deg P-1)^2\partial_\gamma P^{\gamma(\alpha_1}P^{\alpha_2\dots\alpha_{I\min1}}P^{\alpha_I)\mu}\,.\nonumber
\end{eqnarray}
\item The linearity of the first Poisson bracket relation in $L$ allows for a power series ansatz
$$L(x)[\hat{P}](K) = \sum_{i=0}^\infty G(x)(\hat{P}, \partial\hat{P}, \partial\partial\hat{P},\dots )_{A_1\dots A_i} K^{A_1}(x) \dots K^{A_i}(x)\,,$$
so that the problem of finding the local superhamiltonian, and thus the gravitational dynamics, reduces to solving the equations for the coefficients $G_{A_1 \dots A_N}$ that result from making this ansatz, and the corresponding equations are the subject of the following theorem.
\end{enumerate}

\sep{\bf Theorem.} The coefficients $G_{A}, G_{AB}, G_{ABC}, \dots $ defining the local part of the superhamiltonian for the geometry seen by point particles (and thus in conjunction with the already known non-local part and the supermomentum the entire gravitational dynamics) 
\begin{enumerate}
\item[(a)] \underline{depend at most} on the geometric variables $\hat P^A$ and their partial derivatives up to (and including) second order 
\item[(b)] are \underline{completely determined} by the \underline{representation of the deformation algebra} equations
\begin{eqnarray}
\textrm{(I)} \quad \,\,  0 = & & 2 \partial_\mu(G_A R^{A\beta\mu}) +2 G_A S^{A\beta} - 2 \partial_\mu\left(\frac{\partial G}{\partial\partial_{(\mu|}\hat{P}^A} M^{A|\beta)}\right) - 4 \partial_\mu\left(\frac{\partial G}{\partial\partial^2_{(\mu|\nu} \hat{P}^A} \partial_{\nu}M^{A|\beta)}\right)\nonumber \\
   & & + 2 M^{A\beta}\frac{\partial G}{\partial \hat{P}^A} + 2\partial_\mu M^{A\beta} \frac{\partial G}{\partial\partial_{\mu}\hat{P}^A} + 2 \partial^2_{\mu\nu} M^{A\beta} \frac{\partial G}{\partial\partial^2_{\mu\nu}\hat{P}^A}\,,\nonumber\\
\textrm{(II)} \,\quad  0 = & & (N+1)!\, G_{AB_1\dots B_N} R^{A\alpha\beta } - N!\, \frac{\partial G_{B_1\dots B_N}}{\partial\partial_{(\beta|} \hat{P}^A} M^{A|\alpha)} - 2 N!\, \frac{\partial G_{B_1\dots B_N}}{\partial\partial^2_{(\beta|\gamma}\hat{P}^A} \partial_\gamma M^{A|\alpha)}  \nonumber\\
  & & -(N-2) (N-1)! \frac{\partial G_{B_1 \dots B_{N-1}}}{\partial\partial^2_{\alpha\beta} \hat{P}^{B_N}}\,,\nonumber\\  
\textrm{(III)}\quad  0 = & & (N+1)! G_{A B_1 \dots B_N} S^{A\alpha } 
           + (N-1)! \sum_{a=1}^N \frac{\partial G_{B_1 \dots \widetilde{B_a} \dots B_N}}{\partial\partial_{\alpha} \hat{P}^{B_a}} 
           - 2(N-1)! \partial_\gamma  \frac{\partial G_{B_1 \dots B_{N-1}}}{\partial\partial^2_{\alpha\gamma} \hat{P}^{B_N}}\nonumber\\
   & & + N! \frac{G_{B_1\dots B_N}}{\partial \hat{P}^A} M^{A \alpha}        
       + N! \frac{\partial G_{B_1\dots B_N}}{\partial\partial_{\gamma} \hat{P}^A} \partial_\gamma M^{A\alpha} 
       + N! \frac{\partial G_{B_1 \dots B_N}}{\partial\partial^2_{\gamma\delta} \hat{P}^{A}} \partial^2_{\gamma\delta} M^{A\alpha}\,\nonumber\\
   & &   - NN!Q_{(B_1}{}^{M\alpha}G_{B_2\dots B_N)M}\,,\nonumber
\end{eqnarray}
which also include the two \underline{symmetry conditions}
\begin{eqnarray}
\textrm{(IV)} \qquad\qquad\qquad\qquad \frac{\partial G_{B_1 \dots \widetilde{B_a} \dots B_N}}{\partial \partial^2_{\gamma_1\gamma_2}\hat{P}^{B_a}} &=& \frac{\partial G_{B_1 \dots \dots B_{N-1}}}{\partial \partial^2_{\gamma_1\gamma_2}\hat{P}^{B_N}} \qquad \textrm{for all } N\geq 1, \, a = 1,\dots, N\nonumber\\
\textrm{(V)} \,\,\,\quad\qquad\qquad\qquad\qquad\qquad\qquad  0 &=& \frac{\partial G_{B_1 \dots B_N}}{\partial\partial^2_{(\alpha\beta|} \hat{P}^A} M^{A|\gamma)}\qquad \textrm{for all } N\geq 0\,,\nonumber
\end{eqnarray}
as well as the three \underline{invariance conditions} \enlargethispage{2cm}
\begin{eqnarray}\label{invarone}
\textrm{(VI)}\quad 0&=&\sum_{I=2}^{\deg P}~I~\hat P^{\alpha_2\dots\alpha_I (\sigma}\frac{\partial G_{B_1\dots B_N}}{\partial\partial^2_{\mu\nu)} \hat P^{\alpha_2\dots\alpha_I\rho}}\,,\nonumber\\
\textrm{(VII)}\,\,\,\, 0&=&\sum_{I=2}^{\deg P}\left[I~\hat P^{\alpha_2\dots\alpha_I (\mu}\frac{\partial G_{B_1\dots B_N}}{\partial \partial_{\nu)} \hat P^{\alpha_2\dots\alpha_I\rho}}-\partial_\rho \hat P^{\alpha_1\dots\alpha_I}\frac{\partial G_{B_1\dots B_N}}{\partial\partial^2_{\mu\nu}\hat P^{\alpha_1\dots\alpha_I}}+2I~\partial_{\sigma}\hat P^{\alpha_2\dots\alpha_I(\mu}\frac{\partial G_{B_1\dots B_N}}{\partial\partial^2_{\nu)\sigma}\hat P^{\alpha_2\dots\alpha_I\rho}}\right]\,,\nonumber\\
\textrm{(VIII)}\quad & & \sum_{I=2}^{\deg P}\Big [I~\hat P^{\rho\beta_2\dots\beta_I}\frac{\partial G_{B_1\dots B_N}}{\partial \hat P^{\beta_2\dots\beta_I\mu}}+I~\partial_\gamma\hat P^{\rho\beta_2\dots\beta_I}\frac{\partial G_{B_1\dots B_N}}{\partial \partial_\gamma \hat P^{\beta_2\dots\beta_I\mu}} \nonumber\\
&&-\partial_\mu\hat P^{\beta_1\dots\beta_I}\frac{\partial G_{B_1\dots B_N}}{\partial \partial_\rho \hat P^{\beta_1\dots\beta_I}}+I~\partial_{\gamma\delta}\hat P^{\rho\beta_2\dots\beta_I}\frac{\partial G_{B_1\dots B_N}}{\partial \partial_{\gamma\delta} \hat P^{\beta_2\dots\beta_I\mu}}\nonumber\\
&&- 2\partial_{\mu\gamma}\hat P^{\beta_1\dots\beta_I}\frac{\partial G_{B_1\dots B_N}}{\partial \partial_{\rho\gamma} \hat P^{\beta_1\dots\beta_I}}\Big]\nonumber\\
&= & - \delta^{\rho}_{\mu} G_{B_1\dots B_N}-n_1\delta^{\rho}_{(\beta^{(1)}_1}G_{\beta^{(1)}_2\dots \beta^{(1)}_{n_1})\mu B_2\dots B_N}-\dots -n_NG_{B_1\dots B_{N-1}\mu (\beta^{(N)}_2\dots \beta^{(N)}_{n_N}}\delta^{\rho}_{\beta^{(N)}_1)}\,,\nonumber
\end{eqnarray}
where $n_i$ is the number of small indices contained in the capital index $B_i=\beta^{(i)}_1\dots \beta^{(i)}_{n_i}$.
\end{enumerate}

\sep{\bf Remarks.}
\begin{enumerate}
  \item Solving this set of \underline{linear partial differential equations} and consequently solving for the canonical momenta $\hat\pi_A$ in terms of their Legendre duals $\hat K^A$ is the \underline{mathematical} \underline{problem of finding modified gravity theories} for the geometry seen by point particles. For finding dynamics for the geometry seen by fields one proceeds fully analogously. 
  \item For the case $\deg P = 2$ and $\dim M = 4$, it is relatively straightforward to obtain that the above equations determine that the only non-vanishing coefficients are
\begin{eqnarray}
G &=& (2\kappa)^{-1} \sqrt{g}(R-2\lambda)\,,\nonumber\\
G_{\alpha\beta} &=& \rho\sqrt{g}(R_{\alpha\beta}-1/2g_{\alpha\beta} R)+\sigma\sqrt{g}g_{\alpha\beta}\,,\nonumber\\
G_{\alpha\beta \mu\nu} &=&  (16\kappa)^{-1}\sqrt{g} \left[
g_{\alpha \mu}g_{\beta \nu} + g_{\beta \mu}g_{\alpha \nu} - 2 g_{\alpha\beta}g_{\mu\nu}\right]\,,\nonumber
\end{eqnarray}
with the Ricci tensor $R_{\alpha\beta}$ and Ricci scalar $R$ associated with the Lorentzian metric $P^{\alpha\beta} = g^{\alpha\beta}$ and four undetermined real integration constants $\kappa,\lambda,\rho,\sigma$, of which the last two, however, can be set to zero without changing the gravitational equations of motion. The remaining integration constants $\kappa$ and $\lambda$ are the gravitational and cosmological constant, respectively, and must be determined by experiment. This is \underline{Einstein's general relativity}, the simplest gravity theory.
  \item In particular (i), the question whether there are any \underline{modified gravitational dynamics} at all is the question of existence of solutions of the above system for any even $\deg P > 2$; (ii) the question whether there is only one modified gravity theory for each even $\deg P > 2$ amounts to the question of the uniqueness of solutions of the above system; (iii) the question of what the concrete gravitational dynamics are for each $\deg P$, given their existence, is the problem of finding explicit solutions for the above system.
\end{enumerate}

\newpageja
\appendix

\section*{Acknowledgments}
The author thanks Professor Klaus Mecke for the kind invitation to lecture at the inspiring summer school of the Elitestudiengang Physik of the Universities Erlangen-Nuremberg and Regensburg in Obergurgl/Austria for twelve hours during the first week of September 2011, and the participating students and researchers for interesting questions and useful comments.

\section*{Background and further reading}
\noindent In addition to the two research papers,\\[12pt]
\noindent 
{[A]} D. R\"atzel, S. Rivera \& F. P. Schuller, Geometry of physical dispersion relations,
Phys Rev D 2010\\
{[B]} K. Giesel, F. P. Schuller, C. Witte \& M. Wohlfarth, Dynamics of tensorial spacetime geometries, in preparation, \\[12pt]
on which these lectures are based, the following ten textbooks provide the necessary specialist background underlying the results presented in these papers and the present lectures:\\[12pt]
\noindent  
{[1]} C. J. Isham, Differential Geometry for Physicists, 2nd edition\\
{[2]} J. M. Stewart, Advanced General Relativity, Cambridge\\
{[3]} L. H\"ormander, Linear partial differential operators, fourth printing, Springer\\
{[4]} F. W. Hehl and Y. N. Obukhov, Foundations of classical electrodynamics, Birkh\"auser\\
{[5]} J. K. Beem, P. E. Ehrlich, K. L. Easley, Global Lorentzian geometry, 2nd edition, Marcel Dekker\\
{[6]} Z. Shen, Lectures on Finsler geometry, World Scientific 2001\\
{[7]} V. Perlick, Ray optics, Fermat's principle and applications to general relativity, Springer 2000\\
{[8]} B. Hassett, Introduction to algebraic geometry, Cambridge 2007\\
{[9]} K. Sundermeyer, Constrained dynamics (LNP 169), Springer 1982\\
{[10]} R. Rockafellar, Convex analysis, Princeton 1970\\
\end{document}